\documentclass[twocolumn,tighten,times]{aastex631}

\usepackage{amsfonts}
\usepackage{mathrsfs}
\usepackage{bm}
\usepackage{lipsum}
\usepackage{physics}
\usepackage{multirow}
\usepackage{rotating}
\usepackage{graphicx} 
\usepackage{soul}
\usepackage{subfigure}
\usepackage{color}

\newcommand{\feii}{\ifmmode {\rm Fe\ II} \else Fe~{\sc ii}\fi}
\newcommand{\heii}{\ifmmode {\rm He\ II} \else He~{\sc ii}\fi}
\newcommand{\hei}{\ifmmode {\rm He\ I} \else He~{\sc i}\fi}
\newcommand{\oiii}{\ifmmode {\rm~[O\ III]} \else [O~{\sc iii}]\fi}
\newcommand{\nii}{\ifmmode {\rm~[N\ II]} \else [N~{\sc ii}]\fi}
\newcommand{\mgii}{\ifmmode {\rm~Mg\ II} \else Mg~{\sc ii}\fi}
\newcommand{\ciii}{\ifmmode {\rm~C\ III]} \else C~{\sc iii}]\fi}
\newcommand{\civ}{\ifmmode {\rm~C\ IV} \else C~{\sc iv}\fi}
\newcommand{\hb}{\rm{H$\beta$}}
\newcommand{\ha}{\rm{H$\alpha$}}
\newcommand{\hg}{\rm{H$\gamma$}}
\newcommand{\mbh}{\ifmmode {M_{\bullet}} \else $M_{\bullet}$\fi}
\newcommand{\dotm}{\ifmmode {\dot{\mathscr{M}}} \else $\dot{\mathscr{M}}$\fi}
\newcommand{\Rfe}{\ifmmode {{\cal R}_{\rm Fe}} \else ${\cal R}_{\rm Fe}$\fi}

\def\ergscm{\rm erg\,s^{-1}\,cm^{-2}}

\def\kms{\rm km\,s^{-1}}

\shorttitle{Reverberation Mapping of two AGNs }
\shortauthors{Feng et al.}

\begin{document}

\title{\bf \large Reverberation Mapping of Two Variable Active Galactic Nuclei: Probing the Distinct Characteristics of the Inner and Outer Broad-line Regions}

\author[0000-0002-1530-2680]{Hai-Cheng Feng} 
\affiliation{Yunnan Observatories, Chinese Academy of Sciences, Kunming 650216, Yunnan, People's Republic of China}
\affiliation{Key Laboratory for the Structure and Evolution of Celestial Objects, Chinese Academy of Sciences, Kunming 650216, Yunnan, People's Republic of China}
\affiliation{Center for Astronomical Mega-Science, Chinese Academy of Sciences, 20A Datun Road, Chaoyang District, Beijing 100012, People's Republic of China}
\affiliation{Key Laboratory of Radio Astronomy and Technology, Chinese Academy of Sciences, 20A Datun Road, Chaoyang District, Beijing 100101, People's Republic of China}

\author[0000-0003-3823-3419]{Sha-Sha Li}
\affiliation{Yunnan Observatories, Chinese Academy of Sciences, Kunming 650216, Yunnan, People's Republic of China}
\affiliation{Key Laboratory for the Structure and Evolution of Celestial Objects, Chinese Academy of Sciences, Kunming 650216, Yunnan, People's Republic of China}
\affiliation{Center for Astronomical Mega-Science, Chinese Academy of Sciences, 20A Datun Road, Chaoyang District, Beijing 100012, People's Republic of China}
\affiliation{Key Laboratory of Radio Astronomy and Technology, Chinese Academy of Sciences, 20A Datun Road, Chaoyang District, Beijing 100101, People's Republic of China}

\author{J. M. Bai}
\affiliation{Yunnan Observatories, Chinese Academy of Sciences, Kunming 650216, Yunnan, People's Republic of China}
\affiliation{Key Laboratory for the Structure and Evolution of Celestial Objects, Chinese Academy of Sciences, Kunming 650216, Yunnan, People's Republic of China}
\affiliation{Center for Astronomical Mega-Science, Chinese Academy of Sciences, 20A Datun Road, Chaoyang District, Beijing 100012, People's Republic of China}
\affiliation{Key Laboratory of Radio Astronomy and Technology, Chinese Academy of Sciences, 20A Datun Road, Chaoyang District, Beijing 100101, People's Republic of China}

\author[0000-0002-2153-3688]{H. T. Liu}
\affiliation{Yunnan Observatories, Chinese Academy of Sciences, Kunming 650216, Yunnan, People's Republic of China}
\affiliation{Key Laboratory for the Structure and Evolution of Celestial Objects, Chinese Academy of Sciences, Kunming 650216, Yunnan, People's Republic of China}
\affiliation{Center for Astronomical Mega-Science, Chinese Academy of Sciences, 20A Datun Road, Chaoyang District, Beijing 100012, People's Republic of China}

\author[0000-0002-2310-0982]{Kai-Xing Lu}
\affiliation{Yunnan Observatories, Chinese Academy of Sciences, Kunming 650216, Yunnan, People's Republic of China}
\affiliation{Key Laboratory for the Structure and Evolution of Celestial Objects, Chinese Academy of Sciences, Kunming 650216, Yunnan, People's Republic of China}
\affiliation{Center for Astronomical Mega-Science, Chinese Academy of Sciences, 20A Datun Road, Chaoyang District, Beijing 100012, People's Republic of China}

\author[0009-0005-3823-9302]{Yu-Xuan Pang}
\affiliation{Department of Astronomy, School of Physic, Peking University, Beijing 100871, People's Republic of China}
\affiliation{Kavli Institute for Astronomy and Astrophysics, Peking University, Beijing 100871, People's Republic of China}

\author[0000-0002-0771-2153]{Mouyuan Sun}
\affiliation{Department of Astronomy, Xiamen University, Xiamen, Fujian 361005, People’s Republic of China}

\author[0000-0003-4156-3793]{Jian-Guo Wang}
\affiliation{Yunnan Observatories, Chinese Academy of Sciences, Kunming 650216, Yunnan, People's Republic of China}
\affiliation{Key Laboratory for the Structure and Evolution of Celestial Objects, Chinese Academy of Sciences, Kunming 650216, Yunnan, People's Republic of China}
\affiliation{Center for Astronomical Mega-Science, Chinese Academy of Sciences, 20A Datun Road, Chaoyang District, Beijing 100012, People's Republic of China}

\author[0000-0002-2523-5485]{Yerong Xu}
\affiliation{INAF – IASF Palermo, Via U. La Malfa 153, I-90146 Palermo, Italy}
\affiliation{Department of Astronomy \& Physics, Saint Mary's University, 923 Robie Street, Halifax, NS B3H 3C3, Canada}

\author[0009-0000-7791-8192]{Yang-Wei Zhang}
\affiliation{South-Western Institute for Astronomy Research, Yunnan University, Kunming 650500, People’s Republic of China}

\author[0009-0005-2801-6594]{Shuying Zhou}
\affiliation{Department of Astronomy, Xiamen University, Xiamen, Fujian 361005, People’s Republic of China}

\correspondingauthor{Hai-Cheng Feng, Sha-Sha Li}
\email{hcfeng@ynao.ac.cn, lishasha@ynao.ac.cn}

\begin{abstract}
Current reverberation mapping (RM) studies primarily focus on single emission lines, particularly the \hb\ line, which may not fully reveal the geometry and kinematic properties of the broad-line region (BLR). To overcome this limitation, we conducted multiline RM observations on two highly variable active galactic nuclei (AGNs), KUG 1141+371 and UGC 3374, using the Lijiang 2.4 m telescope. Our goal was to investigate the detailed structure of different regions within the BLR. We measured the time lags of multiple broad emission lines (\ha, \hb, \hg, \hei, and \heii) and found clear evidence of radial ionization stratification in the BLRs of both AGNs. Velocity-resolved RM analysis revealed distinct geometry and kinematics between the inner and outer regions of the BLRs. Assuming that velocity-resolved lags reflect the kinematics of BLR, our observations indicate that: (1) in KUG 1141+371, the inner BLR exhibits outflow signatures, while the outer region is consistent with virialized motion; (2) in UGC 3374, the inner region displays virial motion, while the outer region shows inflow. Furthermore, we detected ``breathing" behavior in the outer BLR regions of both AGN, while the inner BLR regions show ``anti-breathing", which may be linked to intrinsic BLR properties. We discuss these findings in the context of various BLR formation models, highlighting importance of long-term, multiline RM campaigns in understanding of BLR structure and evolution. Additionally, our results suggest that the observed stratification in BLR geometry and kinematics may contribute to the scatter in black hole mass estimates and the rapid changes in velocity-resolved RM signatures reported in recent studies.
\end{abstract}


\section{Introduction} \label{sec:intro}
The broad-line region (BLR) in an active galactic nucleus \citep[AGN;][]{Davidson1979, Osterbrock1986, Netzer2013} represents gaseous clouds located deep within the gravitational potential of the central supermassive black hole (SMBH). These high-velocity gaseous clouds are photoionized by the extreme ultraviolet (UV) and soft X-ray emission from the SMBH accretion disk, emitting Doppler-broadened emission lines (BELs) with typical velocities of thousands to several tens of thousands of km s$^{-1}$ \citep{Peterson1997, Netzer2015}. The BLR is generally characterized by an extended geometry, which may include substructures that shape multiple-peaked line profiles \citep{Storchi-Bergmann2017, Du2023}, and diverse kinematics, such as Keplerian motions or inflows governed by the SMBH's gravity \citep{Peterson1999, Peterson2004}, outflows driven by radiation pressure \citep{Marconi2008, Netzer2010, Liu2017, Liu2024}, or possibly turbulence \citep{Osterbrock1978}. Diagnosing these complex structure and kinematics are crucial for probing the gaseous environment around active SMBHs \citep[e.g.,][]{Murray1995, Proga2004, Czerny2011, Gaskell2013, Gaskell2016, Wang2017, Naddaf2021} and for accurately measuring the masses of SMBHs \citep[e.g.,][]{Pancoast2011,Goad2012}.

The diversity of BLR structure and kinematics encodes a complex origin and evolution of these gas clouds. Indeed, several models have been proposed to explain the formation and distribution of BLR clouds, including radiation-driven accretion disk winds \citep{Murray1995, Proga2004}, and clumps originating from the dusty torus that fail to escape the SMBH's gravitational pull \citep{Czerny2011} or undergo tidally disrupted process \citep{Wang2017}. These models predict a complex mixture of inflows, outflows, and virialized motions, as well as diverse geometry, at different radius within the BLR \citep[e.g.,][]{Gaskell2013, Gaskell2016}. Resolving the BLR structure and kinematics is essential for testing these models and unveiling the intricate physical processes governing the BLR.

One powerful technique for resolving the BLR and measuring the SMBH mass (\mbh) is the reverberation mapping (RM) technique \citep{Blandford1982, Peterson1993}. RM estimates the size of the BLR by measuring the response of BELs to variations in the continuum. To date, RM observations have been conducted on over 150 AGNs \citep{Peterson1998, Peterson2004, Kaspi2000, Bentz2009, Hu2021, Malik2023, Woo2024}. Utilizing RM results, a relation between the BLR size ($R$) and the continuum luminosity ($L$) has been established \citep{Kaspi2000, Bentz2013, Du2019}. This relation has also been confirmed by spectroastrometric observations from GRAVITY \citep{GRAVITY2018,GRAVITY2020,Abuter2024}.

With advancements in spectral data quality, the RM technique can now deliver velocity-resolved time lags. This method measures the sizes of the emission regions as a function of velocity within the same BEL and can even recover velocity-delay maps \citep{Bentz2010, Pancoast2011, Li2018, Horne2021}. These techniques have been applied to a growing sample of AGNs, offering detailed constraints on BLR properties \citep[e.g.,][]{DeRosa2018, Lu2019, Williams2021, Li2021, Li2022, Villafa2022, Bao2022, Chen2023}. These studies generally interpret the various velocity-resolved RM measurements as indicative of different BLR kinematic structures. However, \citet{Li2024} found that the ionization distribution of the emission lines across the velocities is inconsistent with the kinematic expectations from velocity-resolved lags, suggesting the necessity for an asymmetric BLR geometry. Moreover, repeated RM observations have revealed that BLRs in some AGNs not only deviate from the expected ``breathing" effect—where $R$ increases with increasing $L$—but also exhibit rapidly changing velocity-resolved signatures that are shorter than dynamical timescales and cannot be explained by radiation pressure acceleration \citep{Cackett2006, Wang2020, Feng2021a, Feng2021b, Feng2024, Lu2022, Yao2024}.

Additionally, the ``Locally Optimally Emitting Clouds" (LOC) model predicts a radially stratified ionization structure within the BLR \citep{Baldwin1995, Korista2000}, where the optimal emitting regions for different emission lines are located at varying distances from the central black hole. This hypothesis has been confirmed by multiline RM observations \citep[e.g.,][]{Bentz2010, Feng2021b, Feng2024}. Consequently, single-line velocity-resolved RM measurements, which are employed in most studies, are insufficient to fully reflect the distribution and velocity field of the entire BLR gas. One possible solution is to incorporate detailed photoionization processes in the measurements, though this approach is computationally expensive \citep{Rosborough2024}. Alternatively, conducting velocity-resolved RM observations on multiple emission lines simultaneously can provide a more comprehensive map of the BLR's geometry and kinematics across different regions.

To better understand the physical properties of the BLR, we carried out multiline RM observations on a sample of AGNs with dramatic luminosity variations. KUG 1141+371 and UGC 3374 are two of our monitoring targets, which have exhibited large-amplitude luminosity variations over the past decade and invariably shown prominent BELs. This is different from another AGN population known as changing-look (CL) AGN, whose spectral types change with luminosity variations, i.e., the appearance or disappearance of BELs \citep{Ricci2023}. This difference motivated us to further explore the physical distinctions between non-CL AGNs and CL AGNs.

The paper is organized as follows. Section~\ref{sec:obs} introduces the long-term light curves of the two targets, as well as observations and data reductions. Section~\ref{sec:measure} describes the measurement of the light curves and their variability characteristics. Section~\ref{sec:analysis} details the measurement of time lags, velocity-resolved time lags, and the determination of \mbh. Section~\ref{sec:discuss} discusses the potential interpretations and implications of our findings. Section~\ref{sec:conclusion} summarizes our conclusions. Throughout this paper, we adopt a cosmology with $H_0 = 67$ km s$^{-1}$ Mpc$^{-1}$, $\Omega_m = 0.32$, and $\Omega_\Lambda=0.68$ \citep{PlanckCollaboration2020}.

\begin{figure*}[!ht]
\centering 
\includegraphics[scale=0.6]{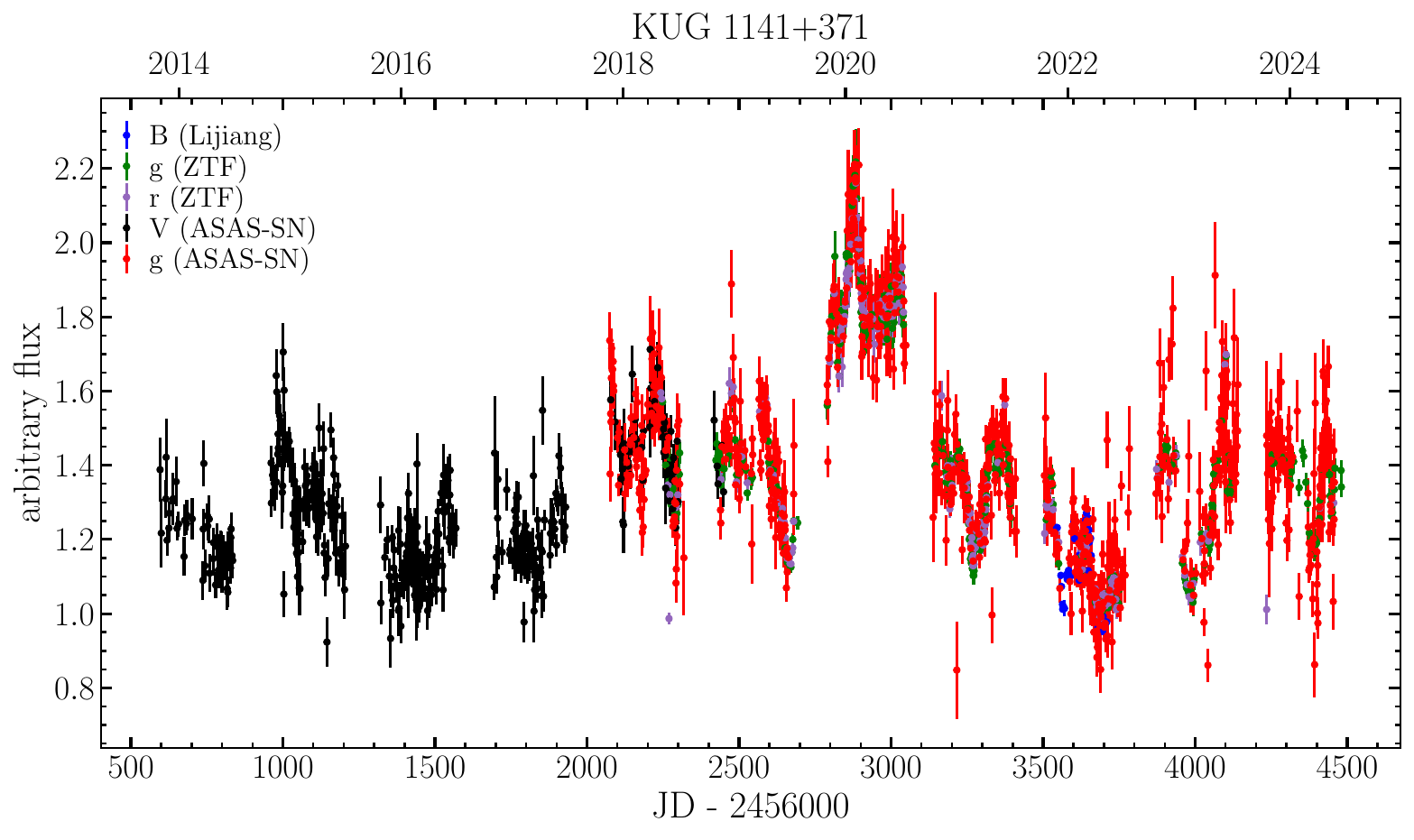}
\includegraphics[scale=0.6]{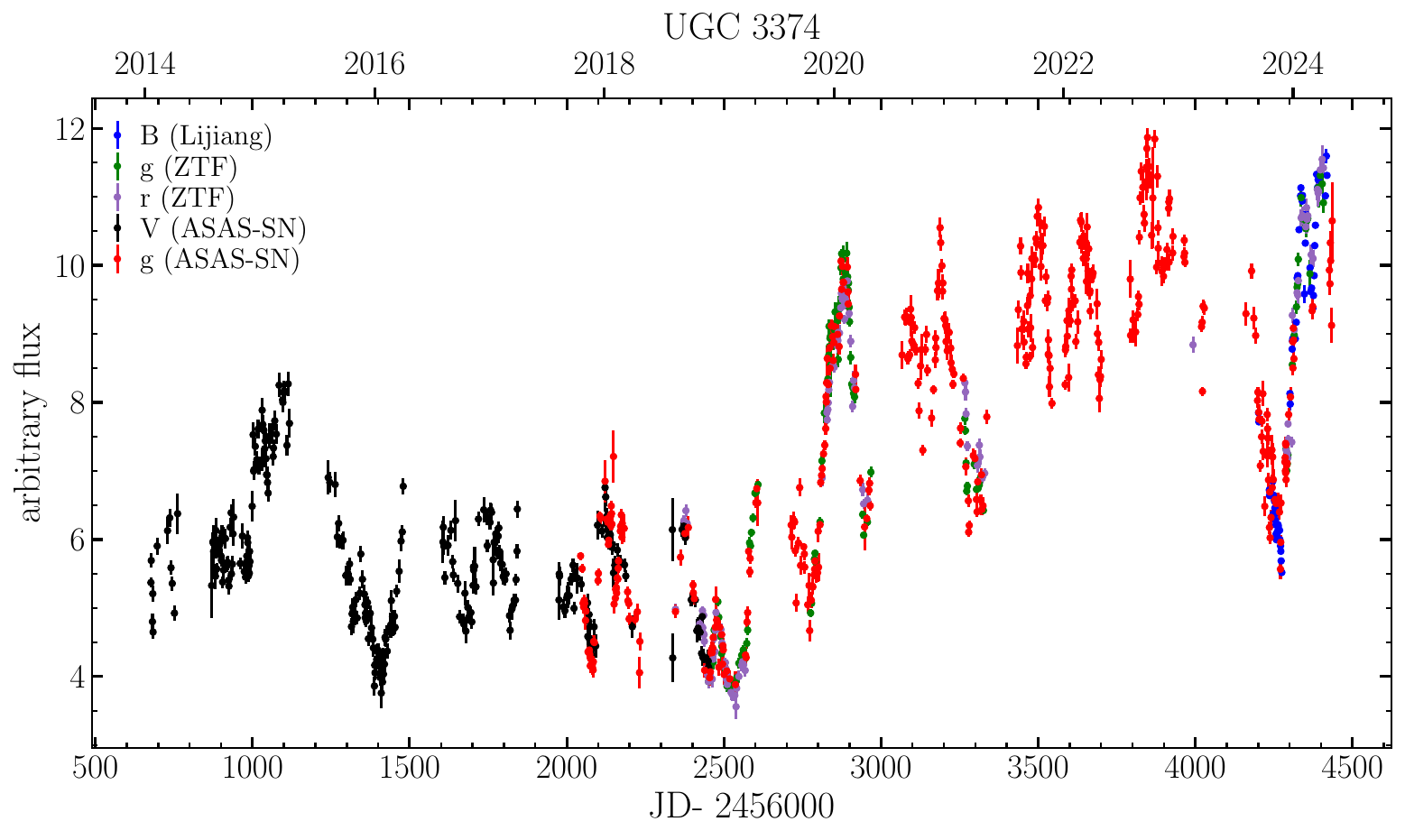}
\caption{Long-term light curves of KUG~1141+371 and UGC~3374. These were collected from multiple sources, including ASAS-SN, ZTF, and our observations at Lijiang. These light curves were calibrated using PyCALI. 
}
\label{fig:longtermlc}
\end{figure*}

\begin{figure*}[!ht]
\centering 
\includegraphics[scale=0.47]{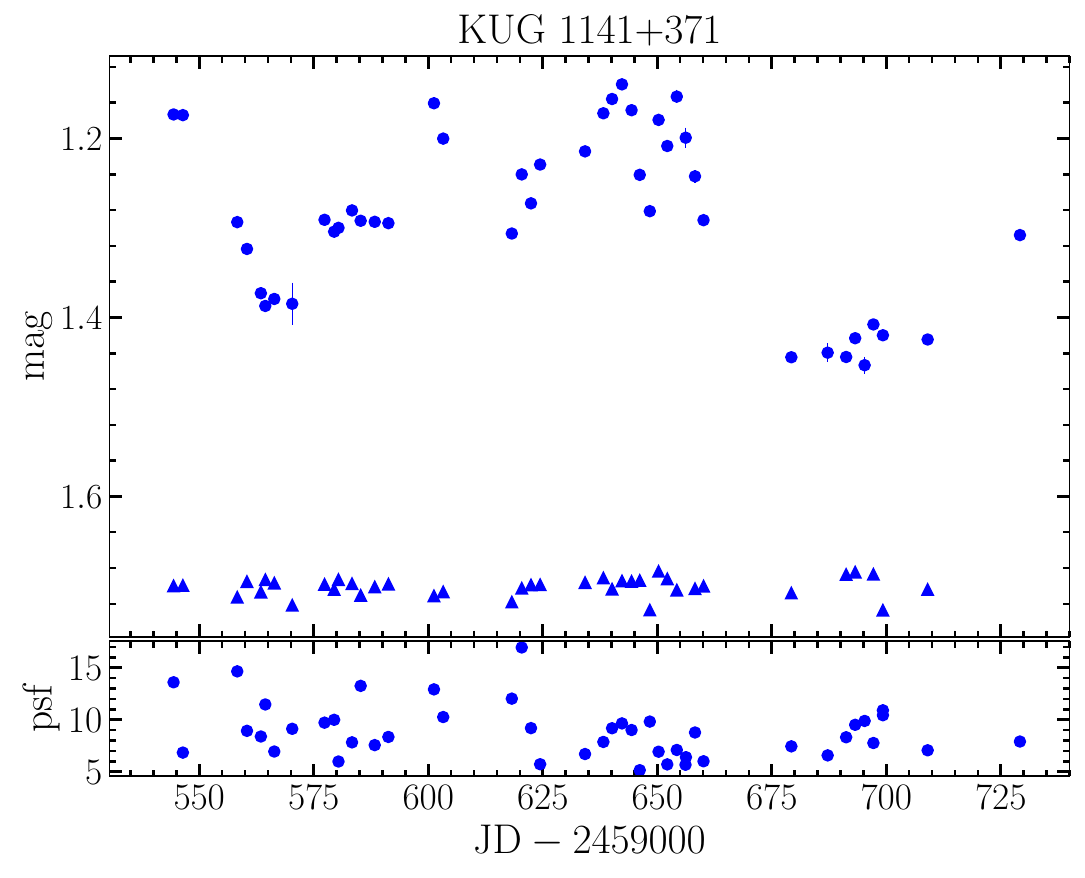}
\includegraphics[scale=0.47]{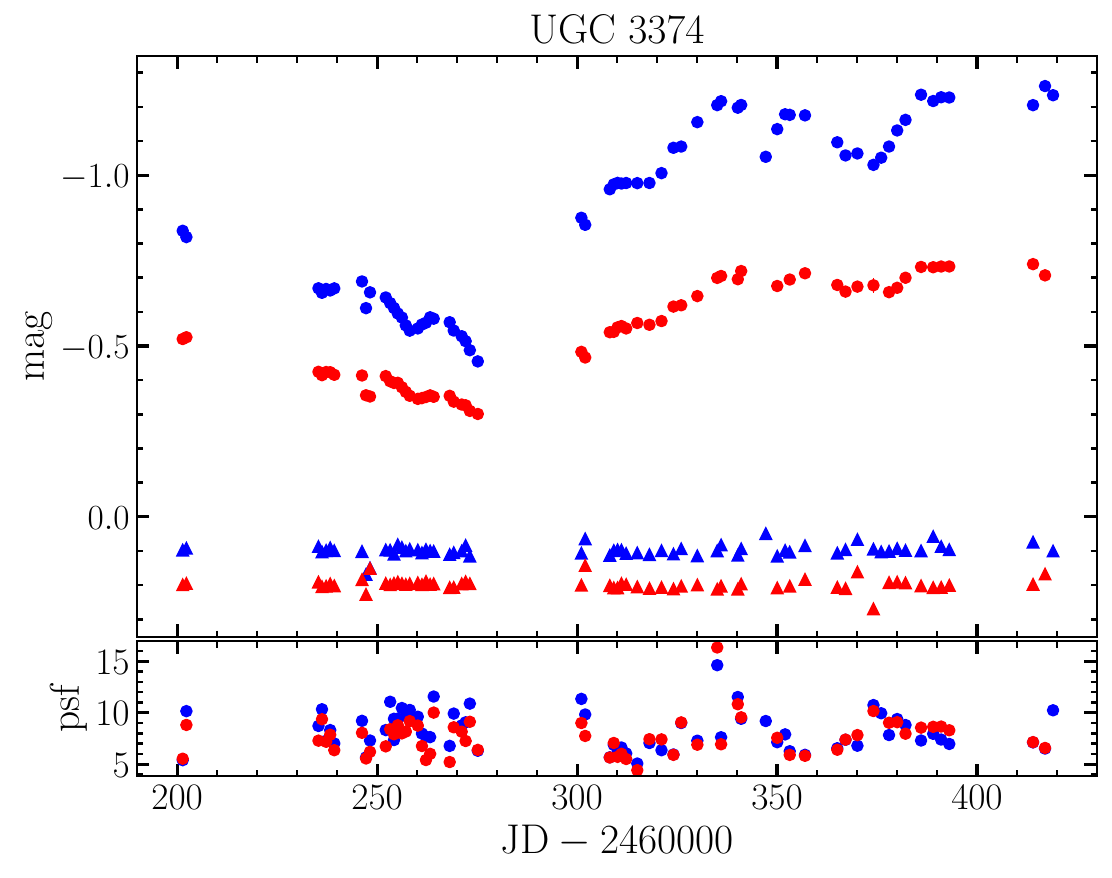}
\caption{Photometric light curves. The left-top panel displays the light curves of KUG 1141+371 and its comparison star, while the right-top panel shows the light curves for UGC~3374 and its comparison star. The dots represent the light curves of the targets, and the upward triangles depict the light curves of comparison stars. Blue represents the $B$ band and red corresponds to the $R$ band. In the bottom of each panel, is the seeing variations, measured as the mean FWHM of stars within the FoV and expressed in pixels.}

\label{fig:photlc}
\end{figure*}

\section{Observations and Data Reduction} \label{sec:obs}
\subsection{Long-term Light Curves from Time-domain Surveys} \label{sec:2.1}
In this study, we analyze long-term light curves from time-domain surveys to identify variable AGNs that are suitable candidates for RM observations. We utilized data from the All-Sky Automated Survey for Supernovae (ASAS-SN) and the Zwicky Transient Facility (ZTF). ASAS-SN\footnote{\url{https://www.astronomy.ohio-state.edu/asassn/index.shtml}} consists of multiple stations equipped with 14cm aperture telescopes, enabling coverage of the entire sky \citep{Shappee2014, Kochanek2017}. The survey provides photometry in the $V$ and $g$ bands, reaching limiting magnitudes of 17.5 and 18.5 mag, respectively. We specifically used light curves in the $V$ and $g$ bands from ASAS-SN Sky Patrol V2.0\footnote{\url{http://asas-sn.ifa.hawaii.edu/skypatrol/}}, which performs aperture photometry following the methodology described by \citet{Hart2023}. 

ZTF\footnote{\url{https://www.ztf.caltech.edu/}} employs a 48inch telescope to survey the northern sky, covering 47 deg$^2$ field of view (FoV) in the $g$, $r$, and $i$ bands. The median depths achieved in these bands are 20.8, 20.6, and 19.9 mag, respectively \citep{Bellm2019, Graham2019, Masci2019}. We obtain ZTF light curves via the Automatic Learning for the Rapid Classification of Events (ALeRCE)\footnote{\url{https://alerce.online/}} broker, which processes and delivers real-time products based on ZTF observations and provides rapid classifications of transient events \citep{Forster2021, Sanchez-Saez2021}. 

To assess the long-term variability properties of the AGNs, we combine data from ASAS-SN, ZTF, and the Lijiang $B$-band photometry (see Section \ref{sec:photo} for details). However, due to the differing contributions from the AGN and galaxy components within these datasets, intercalibration is necessary to ensure consistency. We apply the PyCALI \citep{Pycali2024} software, which adopts the damped random work model to describe variability and Bayesian framework for optimal parameters estimation during the intercalibration process \citep{Li2014}. The intercalibrated light curves are shown in Figure~\ref{fig:longtermlc}, revealing significant variations in KUG~1141+371 and UGC~3374, with a variability amplitude of a factor of 2–3 over the observational baseline. It is worth noting that the host galaxy contamination is not removed in this process, and thus the actual variability amplitude of the AGN should be even larger than observed.

\subsection{Targets} \label{sec:target}
We focus on the RM of two variable AGNs, KUG~1141+371 and UGC 3374. Below, we provide an overview of the key characteristics and previous studies related to these two AGNs.

KUG 1141+371 is a Seyfert 1 galaxy with redshift of 0.038148 and V-band extinction of 0.051 mag \citep{Schlafly2011}. This target has historically exhibited significant variability. Despite the similarity in the shape of X-ray-UV flux ratios and broad-band flux to those observed in CL AGNs, KUG~1141+371 does not exhibit CL characteristics. The AGN remains unobscured across both low and high accretion rate regimes \citep{Jiang2021}, and its spectral type has remained consistent over time \citep{Runco2016}. \cite{Gabanyi2023} identified a compact, flat-spectrum radio feature from VLBI observations, indicating that the radio emissions likely originate from the AGN.  

UGC 3374, also known as MCG +08-11-011, is a bright X-ray and radio-quiet Seyfert galaxy with redshift of 0.020457 and V-band extinction of 0.585 mag \citep{Schlafly2011}. This target has been frequently studied in various RM studies, including optical emission line RM \citep{Fausnaugh2017}, continuum RM \citep{Fausnaugh2018, Fian2023}, and broadband photometric \ha\ RM \citep{Ma2023}. \citet{Fausnaugh2017} conducted optical emission line RM and measured multiple line time lags in UGC 3374, indicting radial stratification of the BLR. Further studies by \citet{Fausnaugh2018} and \citet{Fian2023} in continuum RM revealed that the accretion disk size was larger than predicted by standard disk models. Additionally, \citet{Ma2023} used broadband photometric data from \citet{Fausnaugh2018} and applied the ICCF-Cut method to extract \ha\ light curves and performed analysis. 

\subsection{Photometry} \label{sec:photo}
Our RM project uses the 2.4m telescope at the Lijiang Observatory, Yunnan Observatories, Chinese Academy of Sciences. The telescope is equipped with the Yunnan Faint Object Spectrograph and Camera, which allows for convenient switching between photometry and spectroscopy modes. The FoV is 10$^\prime$ $\times$ 10$^\prime$, enabling the selection of multiple stars within the FoV for photometric calibration. 

We conducted photometric monitoring of two targets. KUG~1141+371 was observed from 2021 November to 2022 May, with 120 s exposures in the $B$ filter over 43 epochs. UGC 3374 was monitored from 2023 September to 2024 April in both the $B$ and $R$ filters, with 65 epochs in the $B$ filter (120 s exposures) and 61 epochs in the $R$ filter (90 s exposures). The photometric data reduction was performed using the standard PyRAF \citep{Pyraf2012} procedures. Following \citet{Li2022}, we employed differential photometry with a circular aperture of 8\farcs49 in diameter (equivalent to 30 pixels). The background annulus was defined with inner and outer radii of 14\farcs15 and 19\farcs81, respectively. Light curves were extracted for both the target and the comparison star, as shown in Figure \ref{fig:photlc}. Our analysis confirmed that the comparison star remained relatively stable, validating its suitability for calibration against the variability of the target.

Considering that seeing variations could cause inconsistent relative contributions from the point source (AGN) and the extended source (host galaxy) within the photometric aperture, potentially increasing the scatter in the light curves \citep{Feng2018}, we further compared light curves extracted using different aperture sizes. We found that the overall variability trends remained nearly identical, with differences at a level of only $\sim$1\%. Additionally, we examined the relationship between seeing and target brightness, and found no correlation (see also Figure \ref{fig:photlc}). This consistency is primarily due to our choice of a relatively large photometric aperture and the relatively low contribution of the host galaxy to the AGN.

\begin{figure*}[!ht]
\centering 
\includegraphics[scale=0.395]{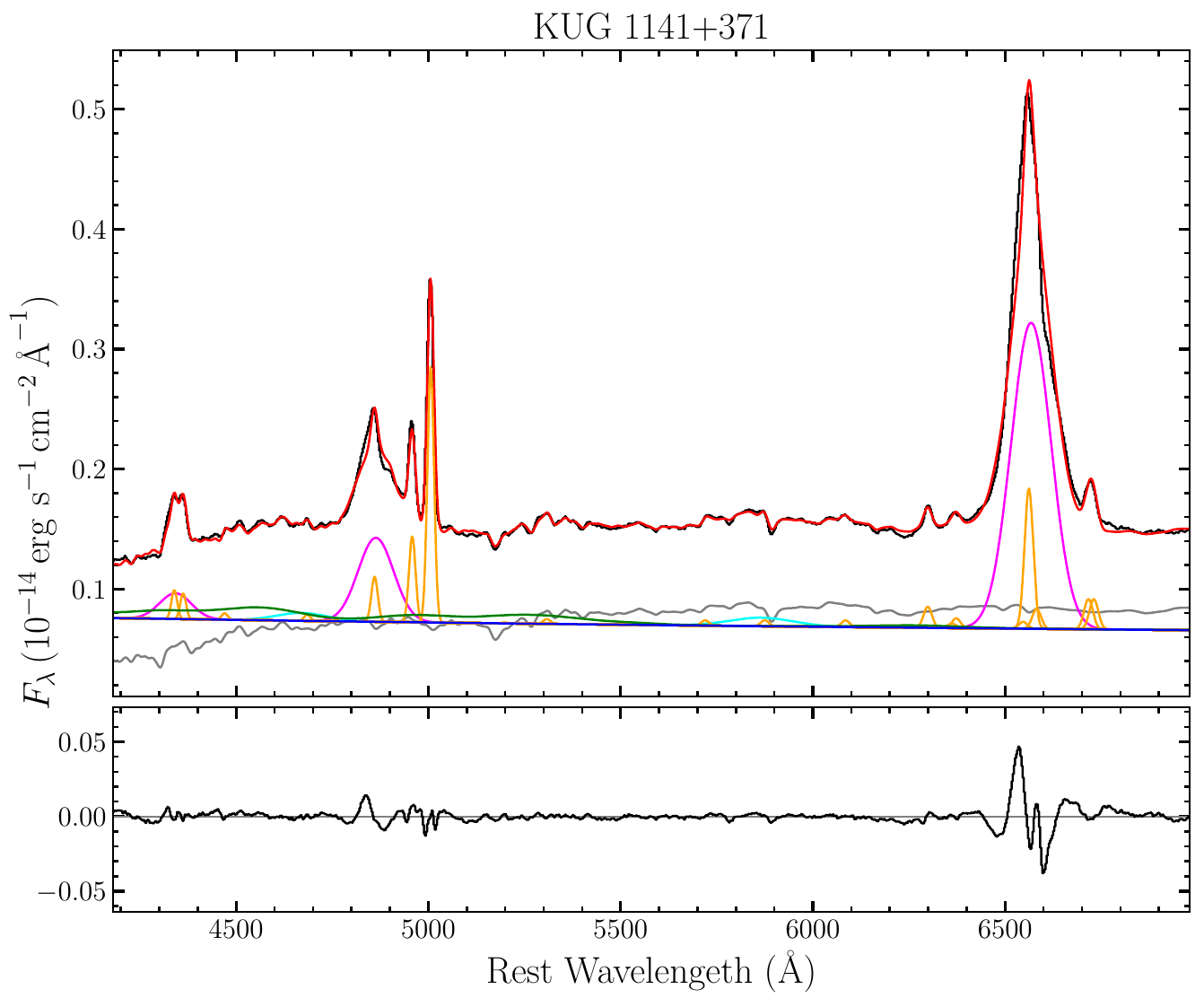}
\includegraphics[scale=0.395]{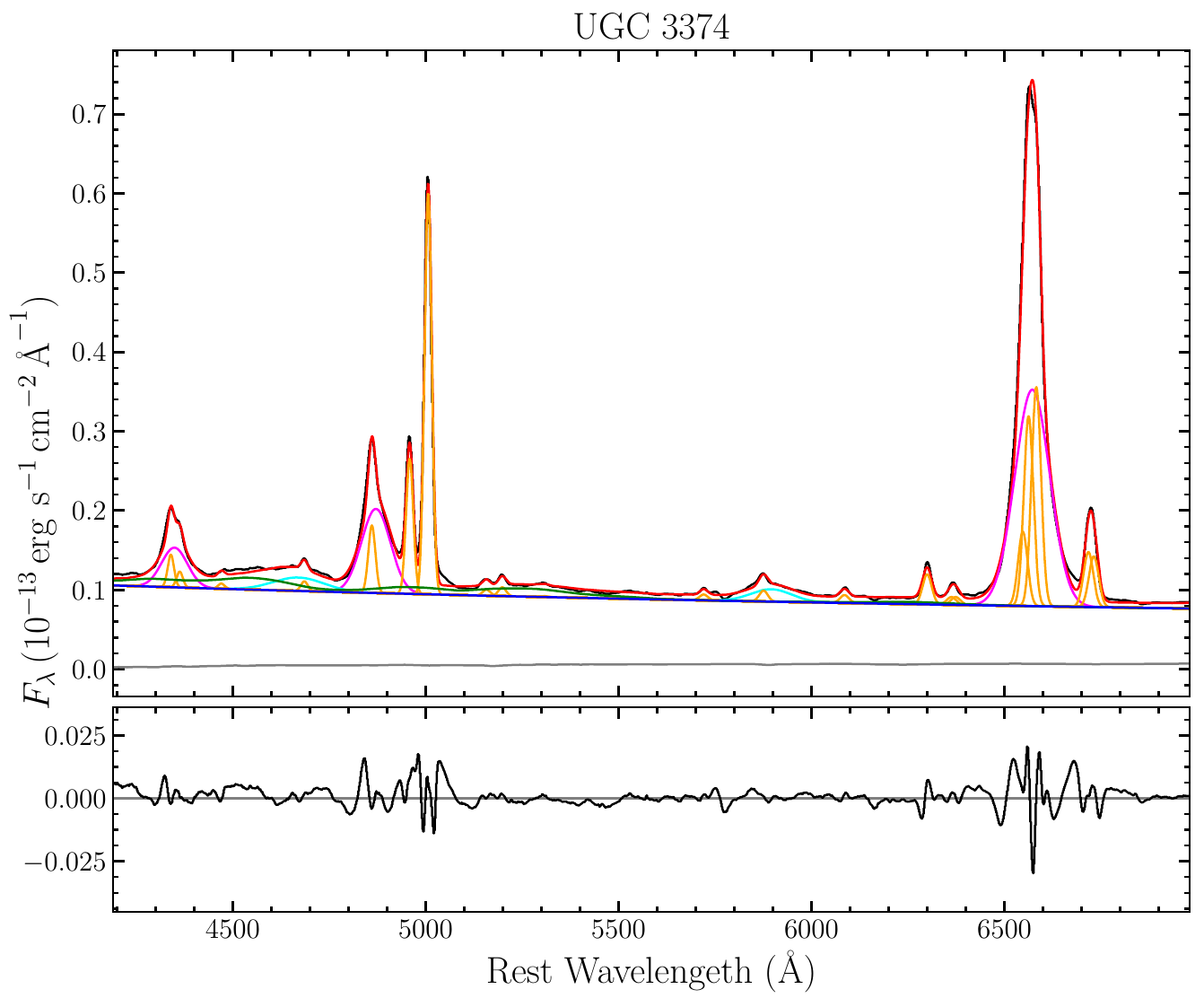}
\caption{Spectral fitting schematic of the mean spectrum for KUG~1141+371 (left) and UGC~3374 (right). Each subplot contains two panels. In the top panel, black line represents the original spectrum, cyan lines depict broad \hei\ and \heii\ lines, magenta lines illustrate broad Balmer lines, orange lines indicate narrow lines, blue line represents the AGN power law, gray line corresponds to the host template, green line shows the \feii\ template, red line represents the best-fitting result. In the bottom panel, black line shows the fitting residuals, and horizontal gray line indicates a value of 0.
}
\label{fig:specfit}
\end{figure*}

\subsection{Spectroscopy} \label{sec:spec}
For spectroscopy observations, we employed a long-slit mode to simultaneously obtain spectra for both the target and comparison star. The advantage of this observational mode is that both objects traverse the same atmospheric conditions, ensuring precise flux calibration and accurate data correction, as described by \citet{Kaspi2000}. For KUG~1141+371, we utilized Grism 3, which provides a dispersion of 2.9 \AA\, pixel$^{-1}$, along with a slit width of 2\farcs5. For UGC~3374, we selected Grism 14, which provides a dispersion of 1.8 \AA\, pixel$^{-1}$, and used a wider slit of 5\farcs05 to reduce light loss. Consistent with our previous studies, a UV-blocking filter was applied to both targets to mitigate secondary spectral contamination, effectively blocking wavelengths shorter than 4150 \AA\ \citep{Feng2020}. The raw data were processed using the PyRAF procedure, aligning the extraction apertures with those used in photometric measurements. The spectral range covered was from 4190 \AA\ to 9050 \AA\ for KUG~1141+371, and from 4190 \AA\ to 7400 \AA\ for UGC~3374. 

Before analysis, the extracted spectra required careful flux calibration involving several key steps:
\begin{enumerate}
\item A standard star observed under optimal weather conditions was selected for the flux calibration of comparison star. The spectra of the flux-calibrated comparison star were then combined to create an initial template spectrum.

\item The preliminary template spectrum of the comparison star was fitted using spectra from the MaNGA Stellar Library. The best-fit spectrum, with the minimal residuals, was selected as the final template. The MaNGA library includes 3321 stars and 8646 spectra from SDSS DR15, covering a wavelength range from 3622 to 10354 \AA\ with a spectral resolution of approximately 1800 \citep{Yan2019}. 

\item The final stellar template spectrum was divided by the nightly observed counts spectrum of the comparison star to generate a response curve. Telluric absorption windows were masked during this process to ensure accuracy.

\item 
The calibrated spectrum of the comparison star was fitted to the stellar template, which accounts for variations in resolution caused by changes in seeing conditions across different nights. To minimize the impact of outliers, data points where the fitting values deviated by less than 6\% from the comparison star spectrum were selected, and the corresponding wavelength information was extracted. At these wavelengths, the template spectrum was divided by the comparison star counts spectrum, again masking telluric absorption windows. This generated a refined response curve for that night, which was used to recalibrate both the comparison star and the target spectra.

\item The final flux-calibrated spectrum of the comparison star was compared to the stellar template to compute the ratio within telluric absorption windows. This ratio was then applied to correct the telluric absorption in the spectrum of the target, resulting in fully flux-calibrated and telluric-corrected spectra for the target.
\end{enumerate}

Each target was exposed for a typical duration of 1500~s, yielding a mean signal-to-noise ratio (S/N) of 34 pixel$^{-1}$ for KUG~1141+371 and 54 pixel$^{-1}$ for UGC~3374 at the rest frame wavelength of 5100 \AA. The analysis included 39 out of 40 spectra for KUG~1141+371 and 57 out of 60 spectra for UGC~3374, ensuring that all evaluated spectra attained an S/N exceeding 10 pixel$^{-1}$. 

\section{Measurements} \label{sec:measure}
\subsection{Light Curve} \label{sec:3.1}
There are two widely used methods to derive emission line light curves from spectral data. The first, commonly referred to as the traditional integration approach, involves selecting two spectral windows around the emission line, estimating the continuum through a linear fit, subtracting it, and then integrating over the emission line region to derive the light curves \citep{Kaspi2000, Bentz2009, Du2018}. However, this method faces challenges when dealing with blended emission lines, such as \heii\ and \feii, which can result in light curves contaminated by contributions from adjacent emission lines. To address this issue, recent studies have employed spectral fitting techniques to isolate individual emission lines, yielding cleaner light curves \citep{Barth2015, Hu2015, Li2021, U2022}. In this work, we adopt the spectral fitting approach to derive the emission line light curves.

Before spectral fitting, we corrected the spectra for Galactic extinction using the extinction curve of \citet{Fitzpatrick1999}, with an assumed value of $R_V = 3.1$. The spectra were then shifted into the rest frame. Additionally, we generate the mean spectrum from individual spectra with S/N $>$ 20 pixel$^{-1}$ to improve the quality of the mean spectrum. 

\begin{deluxetable*}{lcccccccccc}[!htbp]
 \tablecolumns{10}
\tablewidth{\textwidth}
\tabletypesize{\scriptsize}
\tablecaption{Light Curves of KUG~1141+371}
\label{table:lc1}
\tablehead{\multicolumn{6}{c}{Spectra}        &
      \colhead{}                         &
      \multicolumn{3}{c}{Photometry}     \\ 
      \cline{1-6} \cline{8-10}  
      \colhead{JD - 2,459,000}                       &
      \colhead{$F_{\rm H\alpha}$}               &
      \colhead{$F_{\rm H\beta}$}         &
      \colhead{$F_{\rm H\gamma}$}               &
      \colhead{$F_\hei$}               &
      \colhead{$F_\heii$}               &
      \colhead{}                         &
      \colhead{JD - 2,459,000}     &
      \colhead{mag}          &
      \colhead{$\rm Obs$}             
      }
\startdata
544.42 & $39.169 \pm 0.292$ & $10.832 \pm 0.287$ & $2.992 \pm 0.404$ & $1.121 \pm 0.355$ & $1.533 \pm 0.440$ & & 510.01 & $1.089 \pm 0.024$ & ZTF \\
546.44 & $38.623 \pm 0.273$ & $10.425 \pm 0.229$ & $3.229 \pm 0.387$ & $1.650 \pm 0.344$ & $1.388 \pm 0.344$ & & 513.00 & $1.027 \pm 0.026$ & ZTF \\
560.44 & $35.378 \pm 0.265$ & $8.409 \pm 0.223$ & $2.431 \pm 0.382$ & $1.189 \pm 0.334$ & $0.925 \pm 0.341$ & & 522.02 & $0.959 \pm 0.028$ & ZTF \\
564.45 & $33.553 \pm 0.276$ & $7.912 \pm 0.226$ & $2.026 \pm 0.386$ & $1.803 \pm 0.341$ & $0.222 \pm 0.333$ & & 524.05 & $0.976 \pm 0.023$ & ZTF \\
566.41 & $32.680 \pm 0.270$ & $7.385 \pm 0.231$ & $2.443 \pm 0.389$ & $0.720 \pm 0.339$ & $1.033 \pm 0.356$ & & 526.00 & $0.963 \pm 0.023$ & ZTF \\
\enddata
\tablecomments{The emission-line flux is measured in units of $10^{-14}~\ergscm$. The photometry data includes the B-band data from Lijiang and the g-band data from ZTF, which are intercalibrated using PyCALI and are marked with the specific telescope in the ``Obs'' column.
\\
(This table is available in a machine-readable form in the online journal.)}
\end{deluxetable*}

The fitting process was performed using the DAspec \citep{Du2024} software, which employs the Levenberg-Marquardt optimization algorithm. To select an appropriate host galaxy template, we first identified regions in the mean spectrum that were free of prominent emission lines. These regions were then fitted with a combination of an AGN power-law continuum, the \feii\ emission template from \citet{Boroson1992}, and various stellar population templates. We iteratively tested all templates from \citet{Bruzual2003} to identify the best fit, which was adopted as the host galaxy template for subsequent analysis.

Then, we performed a detailed spectral decomposition on both the mean spectrum and individual spectra. The fitting wavelength range spanned from \hg\ to \ha, including several components: (1) an AGN power-law continuum; (2) the optical \feii\ template from \citet{Boroson1992}; (3) the selected host galaxy template from \citet{Bruzual2003}; (4) single Gaussian profiles for broad Balmer lines (\ha, \hb\ and \hg) and Helium lines (\heii\ $\lambda$4686 and \hei\ $\lambda$5876); and (5) single Gaussian for each narrow line. To ensure the robustness of our model for the BELs, we explored alternative models, including double-Gaussians profiles and fourth-order Gauss-Hermite functions. However, the results obtained from these more complex models were consistent with those from the single Gaussian model, which we adopted for simplicity. The velocity profiles of the narrow lines were tied to that of the \oiii\ $\lambda$5007, with fixed flux ratios for the \oiii\ $\lambda\lambda$4959, 5007 and \nii\ $\lambda\lambda$ 6548, 6583 doublets at 1:3 and 1:2.96, respectively. The weaker broad \hg\ emission line, blended with \oiii\ $\lambda$4363 and narrow \hg, was constrained to have the same profile as the broad \hb\ line. By first fitting the mean spectrum, we obtained the flux ratios of all narrow lines relative to \oiii\ $\lambda$5007 and the spectral index of the AGN continuum. These parameters were applied to the fitting of individual spectra. The spectral decomposition results for the mean spectra of KUG1141+371 and UGC3374 are presented in Figure~\ref{fig:specfit}.

\begin{deluxetable*}{lcccccccccc}[!htbp]
 \tablecolumns{10}
\tablewidth{\textwidth}
\tabletypesize{\scriptsize}
\tablecaption{Light Curves of UGC~3374}
\label{table:lc2}
\tablehead{\multicolumn{6}{c}{Spectra}        &
      \colhead{}                         &
      \multicolumn{3}{c}{Photometry}     \\ 
      \cline{1-6} \cline{8-10}  
      \colhead{JD - 2,460,000}                       &
      \colhead{$F_{\rm H\alpha}$}               &
      \colhead{$F_{\rm H\beta}$}         &
      \colhead{$F_{\rm H\gamma}$}               &
      \colhead{$F_\hei$}               &
      \colhead{$F_\heii$}               &
      \colhead{}                         &
      \colhead{JD - 2,460,000}     &
      \colhead{mag}          &
      \colhead{$\rm Obs$}             
      }
\startdata
201.40 & $33.848 \pm 0.228$ & $10.614 \pm 0.172$ & $4.579 \pm 0.168$ & $2.102 \pm 0.058$ & $2.484 \pm 0.225$ & & 161.12 & $-1.029 \pm 0.023$ & ASAS-SN \\
202.32 & $34.091 \pm 0.229$ & $10.817 \pm 0.172$ & $4.500 \pm 0.169$ & $2.160 \pm 0.060$ & $2.826 \pm 0.225$ & & 179.08 & $-1.103 \pm 0.016$ & ASAS-SN \\
235.35 & $30.447 \pm 0.228$ & $9.280 \pm 0.172$ & $3.599 \pm 0.168$ & $1.800 \pm 0.058$ & $1.456 \pm 0.225$ & & 186.06 & $-1.022 \pm 0.023$ & ASAS-SN \\
236.23 & $30.441 \pm 0.228$ & $9.170 \pm 0.172$ & $3.343 \pm 0.169$ & $1.720 \pm 0.059$ & $1.800 \pm 0.226$ & & 192.04 & $-0.990 \pm 0.016$ & ASAS-SN \\
237.27 & $30.558 \pm 0.228$ & $9.239 \pm 0.172$ & $3.777 \pm 0.168$ & $1.900 \pm 0.059$ & $1.744 \pm 0.224$ & & 197.96 & $-0.862 \pm 0.017$ & ASAS-SN \\
\enddata
\tablecomments{The emission-line flux is measured in units of $10^{-13}~\ergscm$. The photometry data includes the B-band data of Lijiang and the g-band data from ZTF and ASAS-SN, which are intercalibrated by PyCALI and are specified with the corresponding telescope in the ``Obs'' column.
\\
(This table is available in a machine-readable form in the online journal.)}
\end{deluxetable*}

\begin{figure*}[!ht]
\centering 
\includegraphics[scale=0.5]{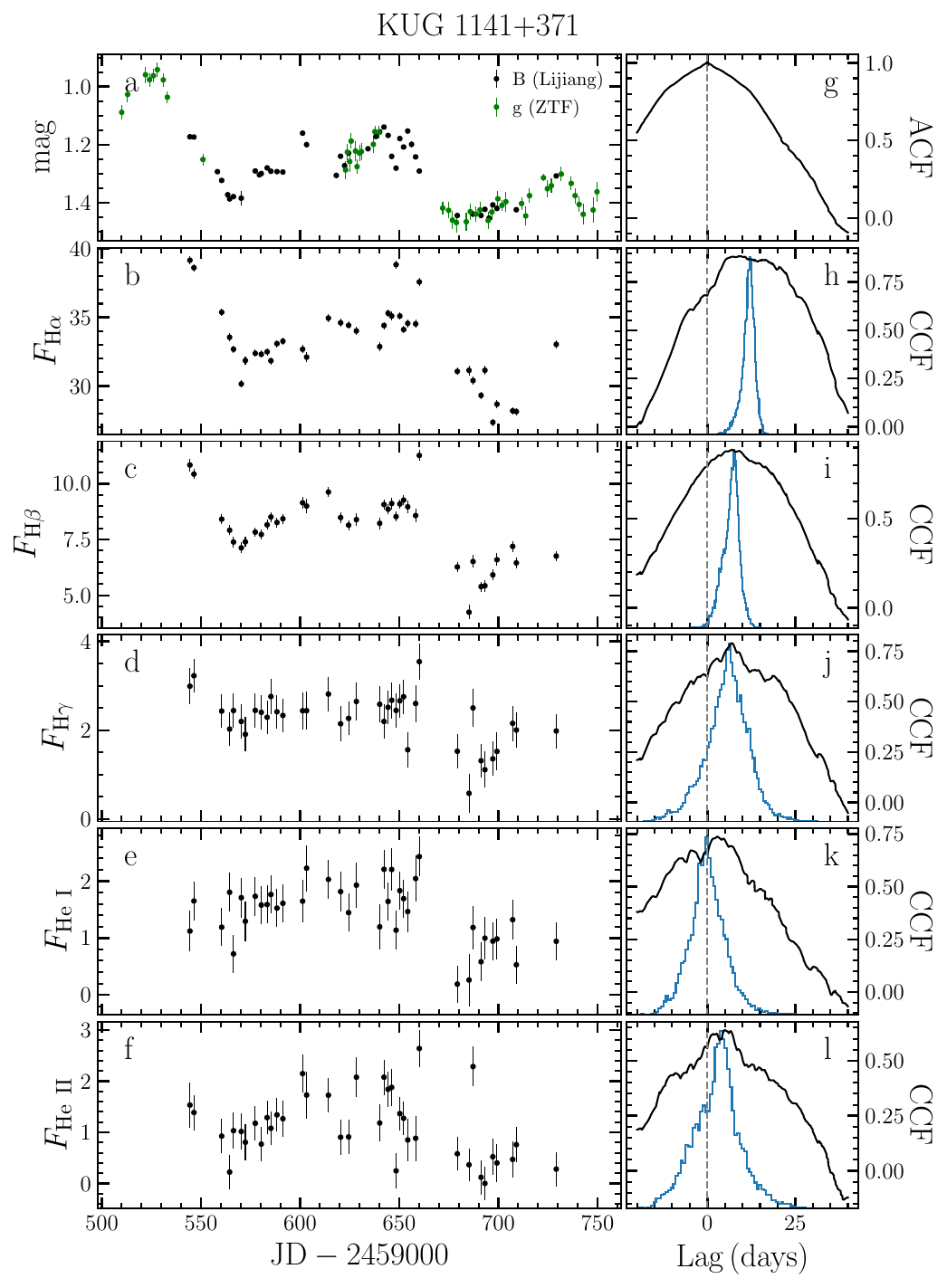}
\includegraphics[scale=0.5]{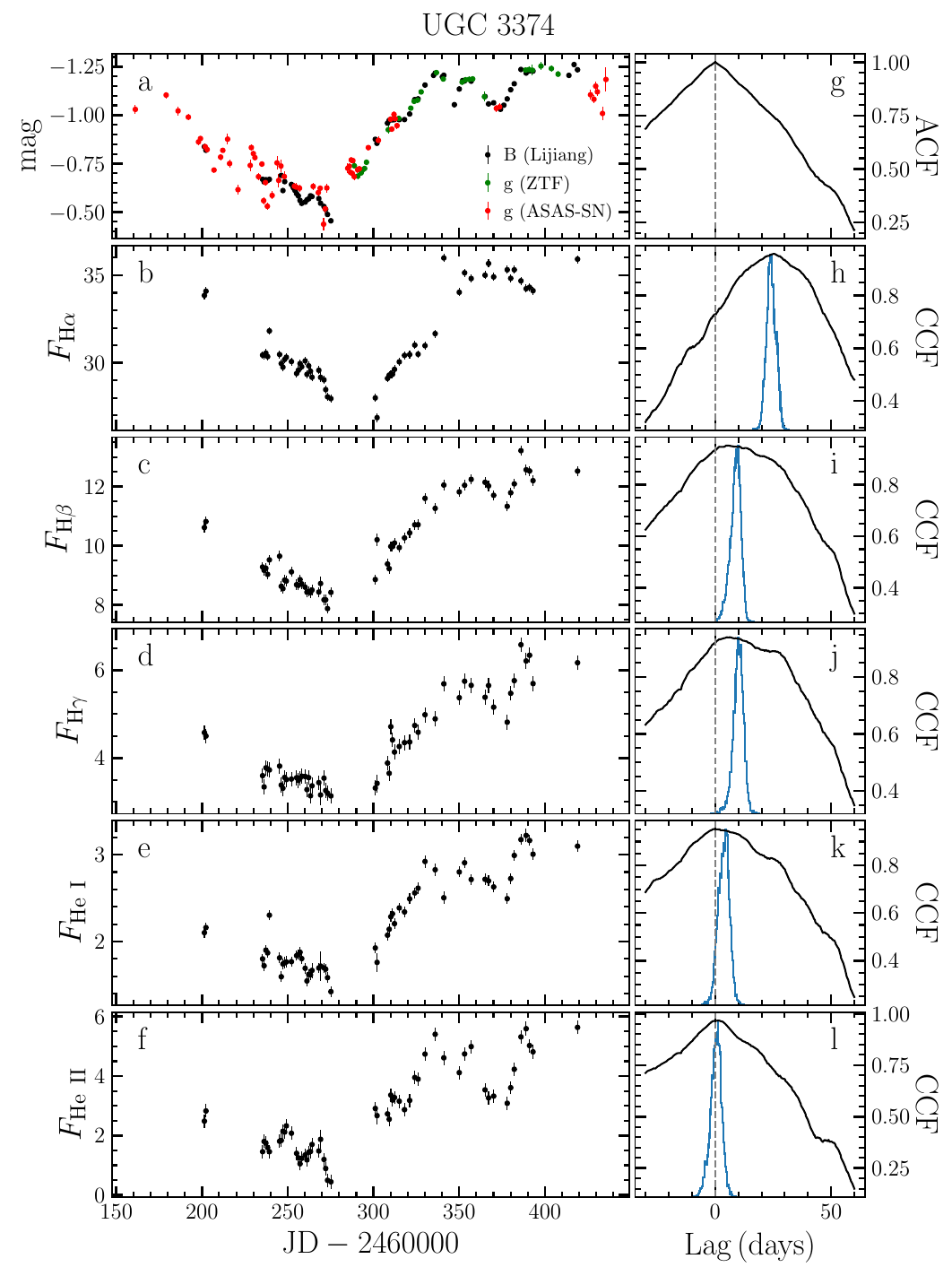}
\caption{Light curves and CCF results for KUG~1141+371 (left) and UGC~3374 (right). Panels (a)-(f) represent light curves for photometry, \ha, \hb, \hg, \hei\ and \heii. The units for emission lines are 10$^{-14} \ergscm$ for KUG~1141+371 and 10$^{-13} \ergscm$ for UGC~3374, respectively. The photometry light curves for KUG~1141+371 are intercalibrated results from B (Lijiang) and g (ZTF). For UGC~3374, the results are intercalibrated from B (Lijiang), g (ZTF), and g (ASAS-SN). Panel (g) displays the ACF result for the photometry light curve. Panels (h)-(l) present the CCF results (black lines) and the CCCDs (blue lines). 
}
\label{fig:speclc}
\end{figure*}

Based on the fitting results, we extracted light curves for the BELs, including the broad Balmer and Helium lines. Observational data can be affected by various external factors, such as weather conditions, target miscentering in the slit, and the impact of instrumental broadening on the fitting. Therefore, we introduced additional uncertainties to account for potential systematics. Following the approach of \citet{Li2022}, we calculated and incorporated these systematic uncertainties into the light curves in quadrature. Additionally, we also used the bootstrap method to remeasure the flux and velocities of the emission lines. The results confirm that our fitting measurements and error estimates are robust (see Appendix \ref{sec:appecdix} for details). The final results are presented in Figure~\ref{fig:speclc}, with detailed results provided in Tables~\ref{table:lc1} and \ref{table:lc2}.

\begin{deluxetable}{lcccccccccc}[!ht]
 \tablecolumns{10}
\tablewidth{\textwidth}
\tabletypesize{\scriptsize}
	\tablecaption{Light Curve Statistics \label{table:stas}}
	\tablewidth{\textwidth}
		\tablehead{\colhead{Object} &
			\colhead{Light Curve} &
			\colhead{Mean Flux} &
			\colhead{$F_{\rm var}(\%)$} &
			\colhead{$R_{\rm max}$}
		}
\startdata
KUG~1141+371 & \ha & $33.09\pm 2.79$ & $8.50\pm0.97$ & 1.37\\
& \hb & $8.02 \pm 1.46$ & $18.19\pm2.12$ & 1.92\\
& \hg & $2.26 \pm 0.57$ & $18.68\pm3.99$ & 2.28\\
& \hei & $1.44 \pm 0.53$ & $28.15\pm5.63$ & 4.17\\
& \heii & $1.11 \pm 0.64$ & $49.13\pm7.93$ & 9.70\\
\hline
UGC~3374 & \ha & $31.27 \pm 2.49$ & $7.99\pm0.75$ & 1.27\\
& \hb & $10.06 \pm 1.49$ & $14.88\pm1.41$ & 1.53\\
& \hg & $4.29 \pm 0.99$ & $22.87\pm2.21$ & 1.96\\
& \hei & $2.20 \pm 0.51$ & $23.37\pm2.23$ & 2.00\\
& \heii & $2.77 \pm 1.41$ & $50.56\pm4.86$ & 6.08\\
\enddata
\tablecomments{The mean fluxes are in units of $10^{-14}~\ergscm$ for KUG~1141+371 and $10^{-13}~\ergscm$ for UGC~3374.}
\end{deluxetable}

\subsection{Variability Characteristics}
We employ $F_{\rm var}$ and $R_{\rm max}$ to characterize the amplitude of variability. Following the definition by \citet{Rodriguez-Pascual1997}, the $F_{\rm var}$ is calculated as 
\begin{equation}
F_{\rm var}=\frac{(S^2-\triangle^2)^{1/2}}{\langle F \rangle},
\end{equation}
where $\langle F \rangle =\frac{1}{N} \sum\limits_{i=0}^N F_{i}$ represents the mean flux, $S^2 = \frac{1}{N-1}\sum\limits_{i=0}^N (F_{i} - \langle F \rangle)^2$ denotes the sample variance, $\triangle^2 = \frac{1}{N}\sum\limits_{i=0}^N \triangle_{i}^2$ is the mean squared error, $N$ is the number of data points, $F_{i}$ is the flux of the $i$-th data point, and $\triangle_{i}$ is the error associated with $F_{i}$. 
The uncertainty in $F_{\rm var}$, as described by \citet{Edelson2002}, can be written as 
\begin{equation}
\sigma_{\rm var}=\frac{1}{F_{\rm var}}\bigg(\frac{1}{2N}\bigg)^{1/2}\frac{S^2}{{\langle F \rangle}^2}.
\end{equation}

The $R_{\rm max}$ is defined as the ratio of the median of the five highest flux values to the median of the five lowest flux values in the dataset. We list the results of $F_{\rm var}$ and $R_{\rm max}$ for each light curve in Table~\ref{table:stas}. Comparing the variability amplitudes across different emission lines, we find that the Balmer lines exhibit lower variability than the Helium lines, consistent with the findings of previous studies \citep[e.g.,][]{Bentz2010}.

\section{Analysis and Results} \label{sec:analysis}
\subsection{Time lags} 
We employed the interpolated cross-correlation function (CCF) method \citep{Gaskell1986, Gaskell1987}, with the modifications introduced by \citet{White1994}, to measure the time lag of emission lines relative to continuum variations. Consistent with our previous studies, we used only photometric data for the continuum light curve.\footnote{This choice was made because the sampling frequency of the spectroscopic and photometric data is similar, and the time delay between the $B$ band and 5100\AA\ continuum is consistent. However, spectroscopic data are more affected by host galaxy contamination, leading to large scatter in the light curves.} To extend the temporal baseline and improve the sampling of the continuum data, we incorporated additional data from time-domain surveys, as described in Section \ref{sec:2.1}. Note that, instead of directly using these data, we performed a re-calibration to enhance the accuracy of the flux measurements. For UGC3374, the continuum light curve was generated by combining the intercalibrated data from the $B$ band (Lijiang), $g$ band (ZTF), and $g$ band (ASAS-SN). In the case of KUG 1141+371, due to the lower S/N of the ASAS-SN data, we used the intercalibrated results from the $B$ band (Lijiang) and $g$ band (ZTF).

\begin{figure*}[!ht]
\centering 
\includegraphics[scale=0.249]{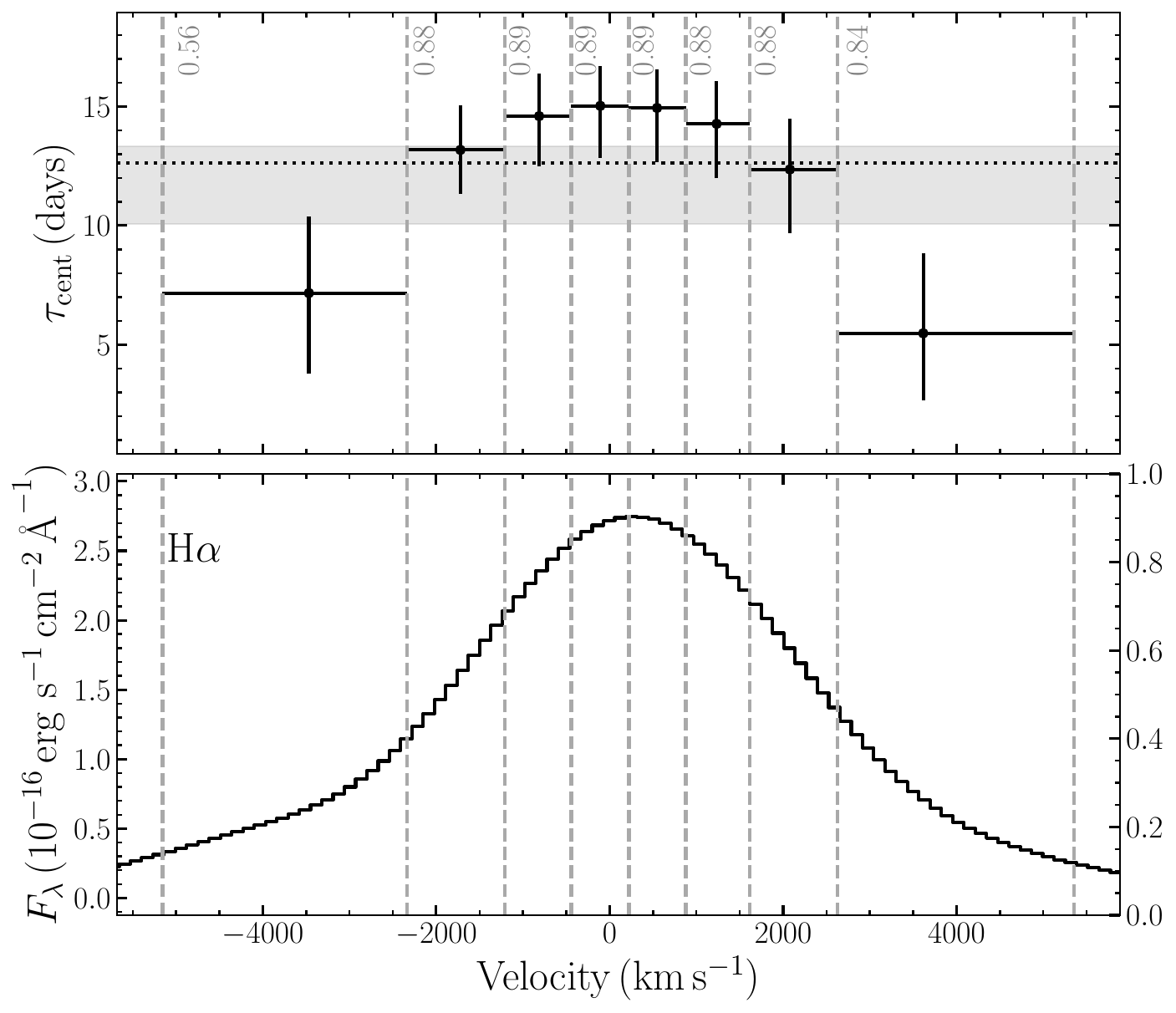}
\includegraphics[scale=0.249]{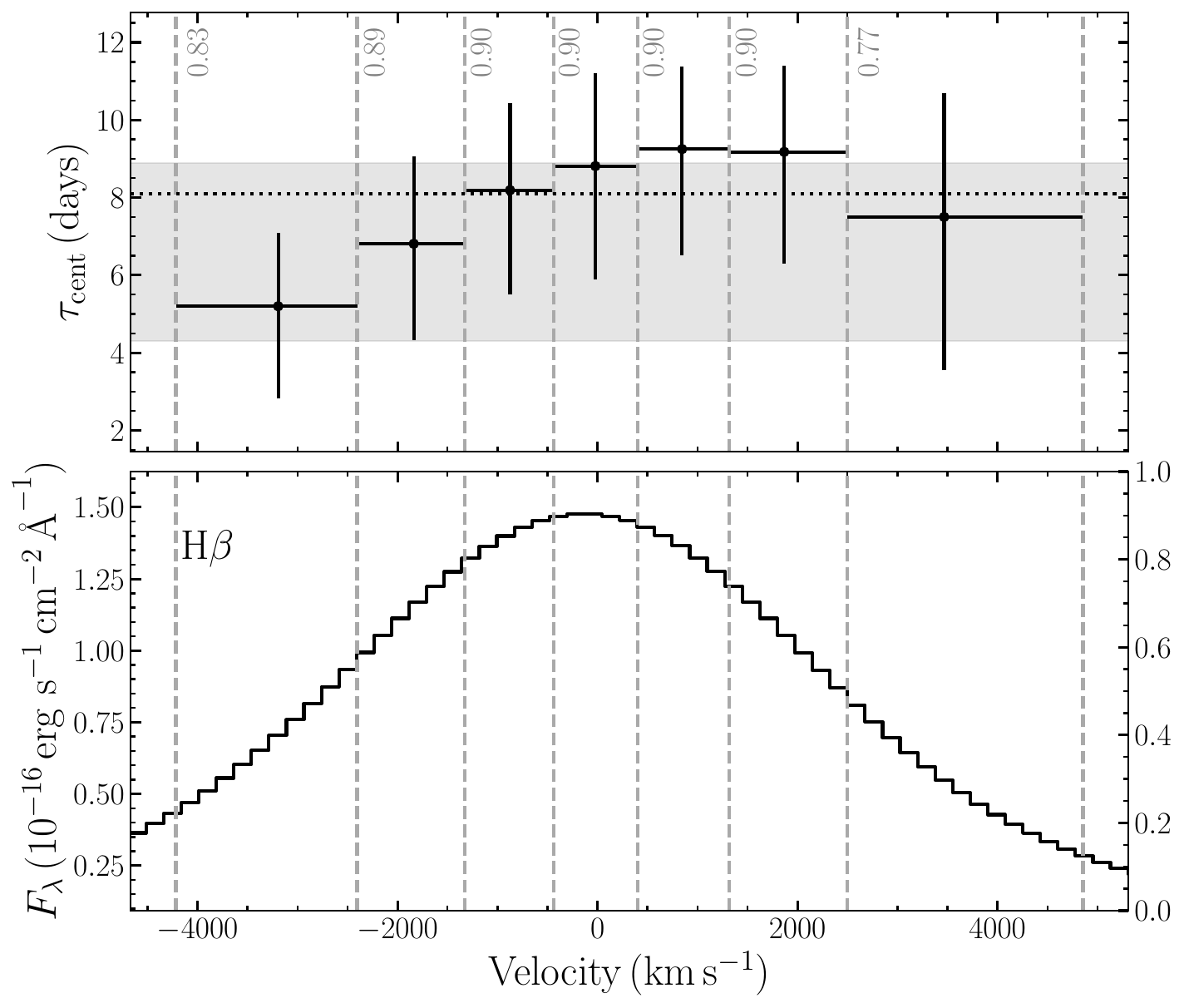}
\includegraphics[scale=0.249]{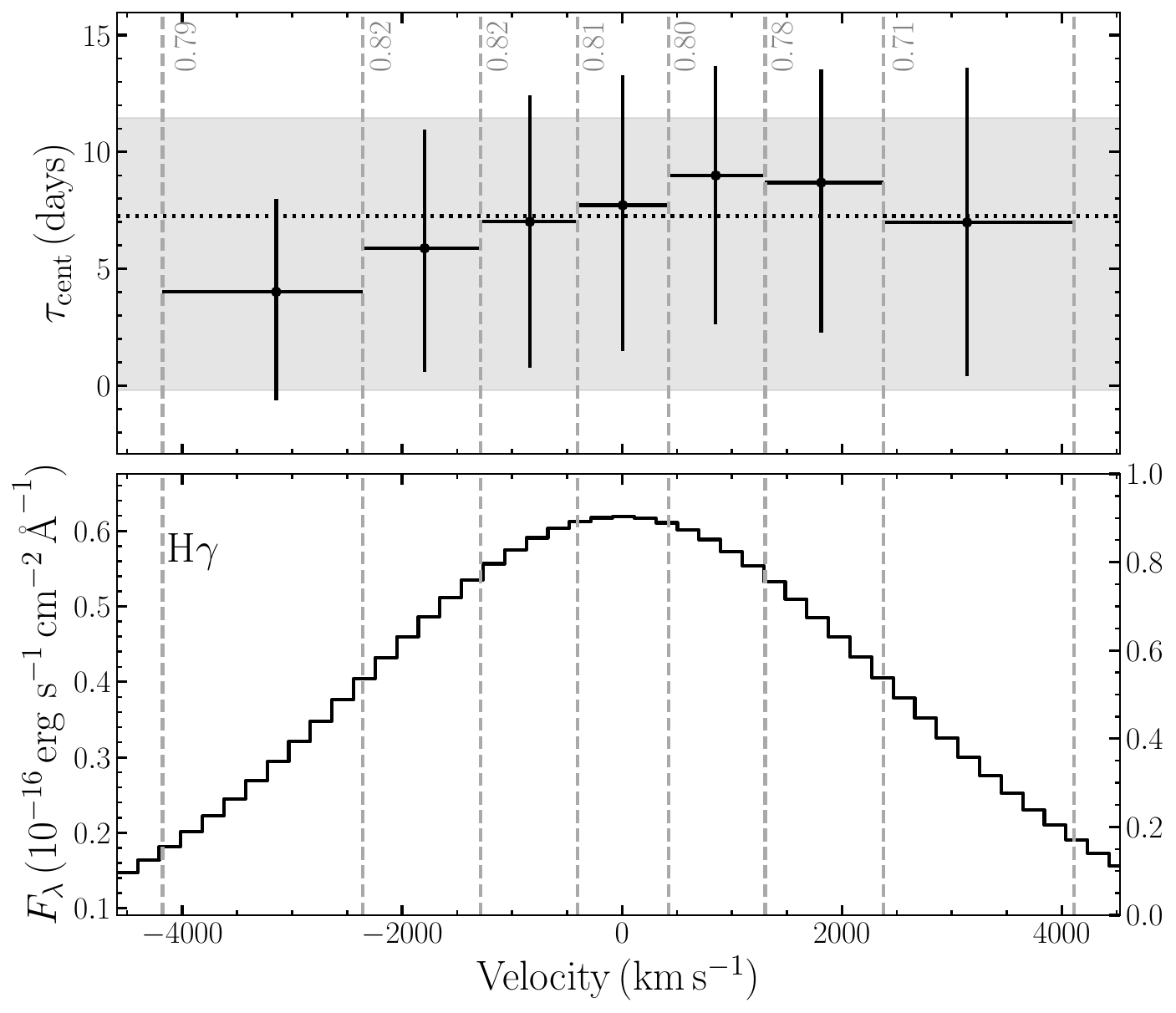}
\includegraphics[scale=0.249]{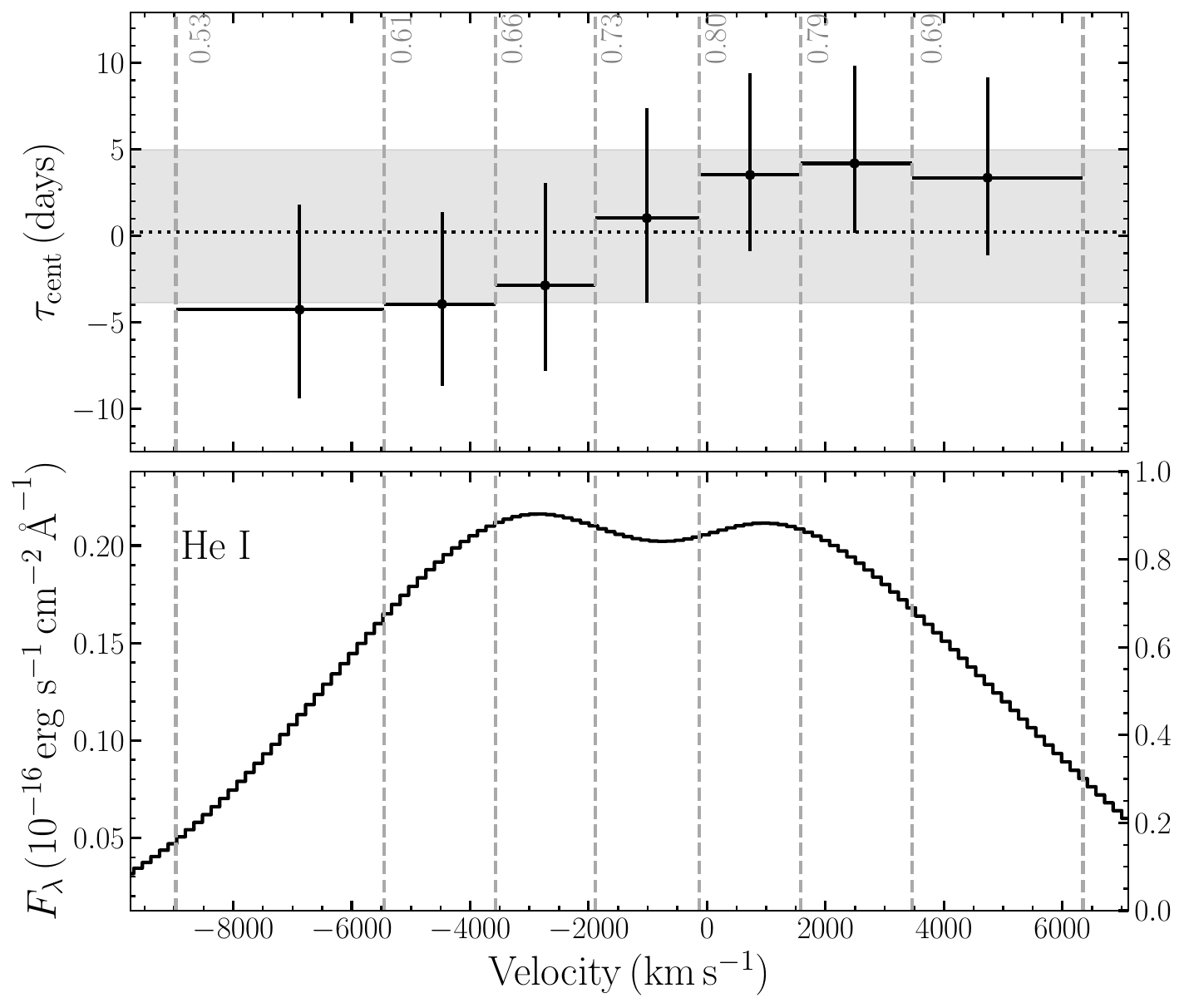}
\includegraphics[scale=0.249]{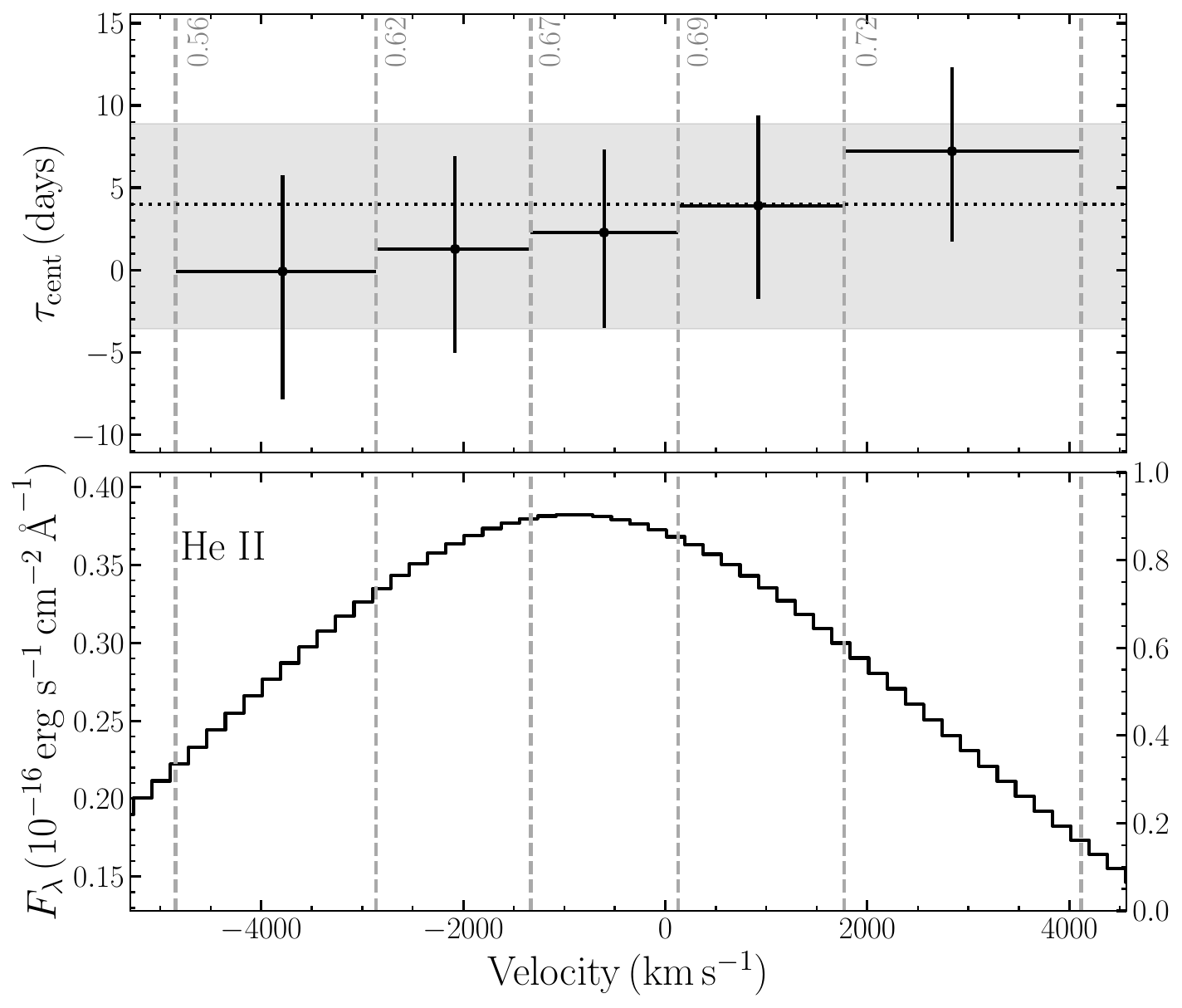}
\caption{Velocity-resolved time delays for KUG~1141+371. Each subplot presents the results for different emission lines and is divided into two sections. The top panel shows the time delay for each velocity bin. The horizontal dotted line represents the mean delay for each emission line, the gray band indicates the corresponding error range, and the vertical gray dashed lines mark the boundaries of each velocity bin. The $r_{\rm max}$ for each bin, measured using the CCF method, is also displayed as a gray numerical value within each bin. In the bottom panel, the black step line illustrates the rms spectrum generated from the profile of each broad emission line, with the vertical gray dashed lines similarly marking the boundaries of each velocity bin. 
}
\label{fig:vr1}
\end{figure*}

\begin{figure*}[!ht]
\centering 
\includegraphics[scale=0.26]{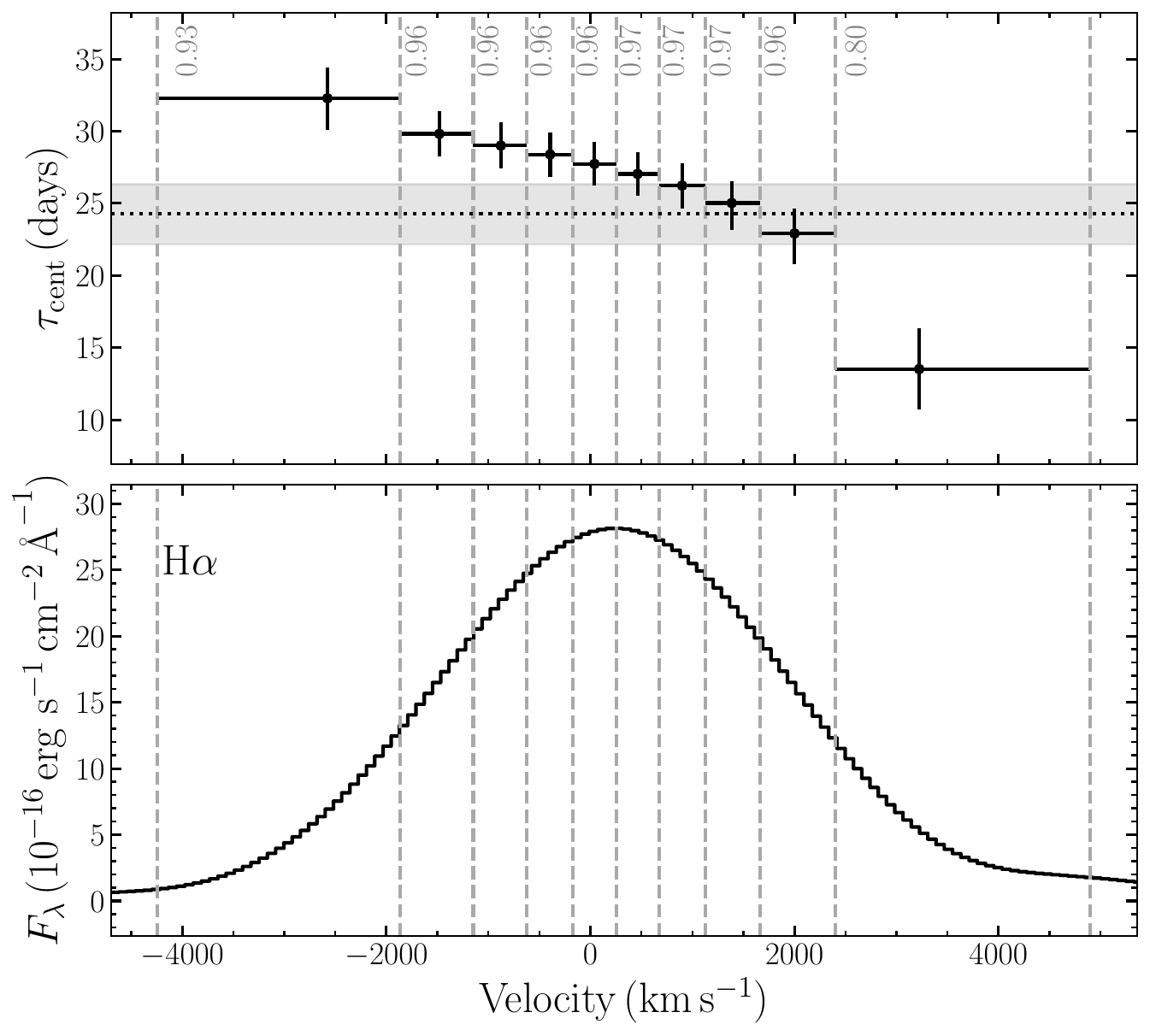}
\includegraphics[scale=0.26]{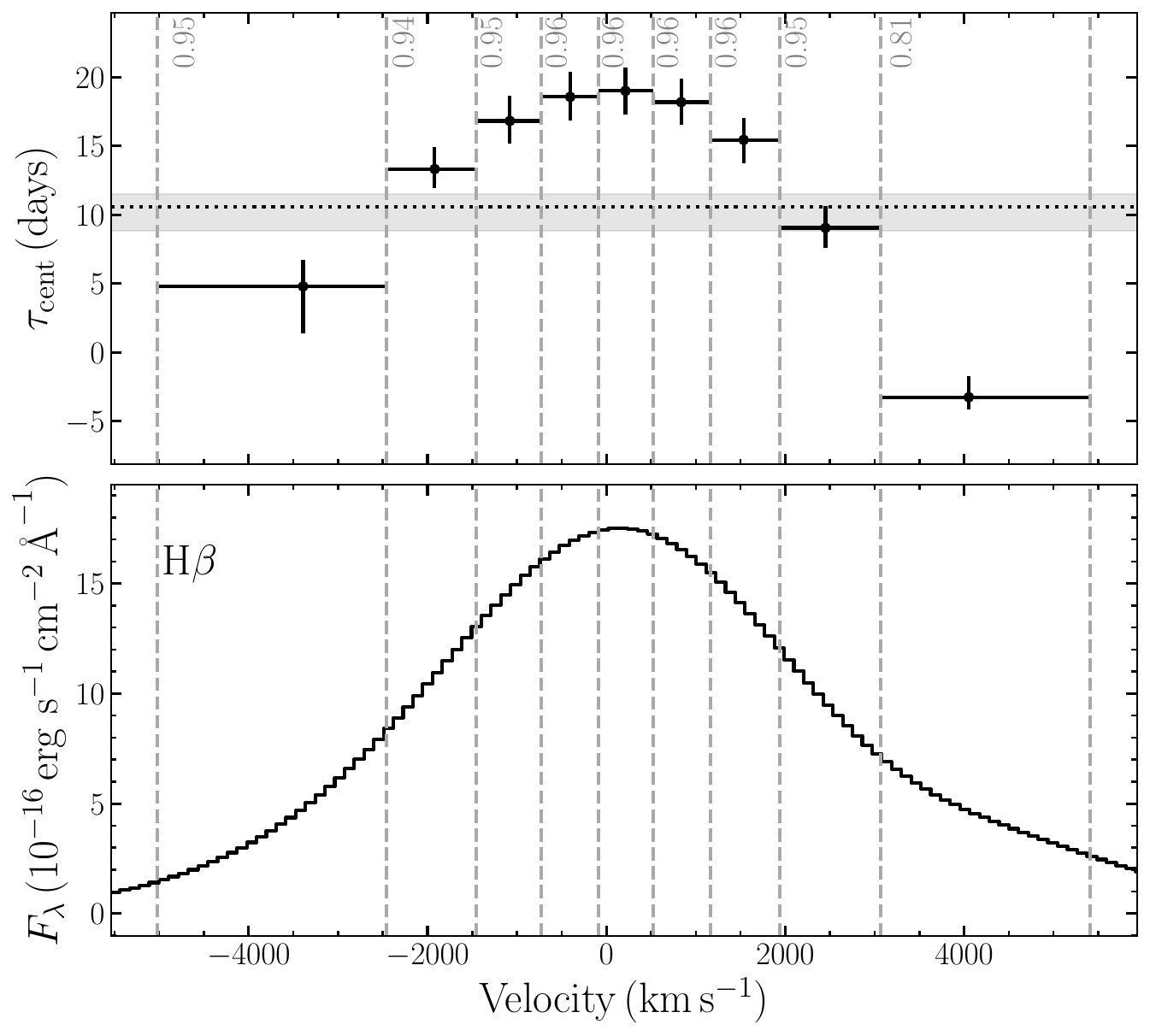}
\includegraphics[scale=0.26]{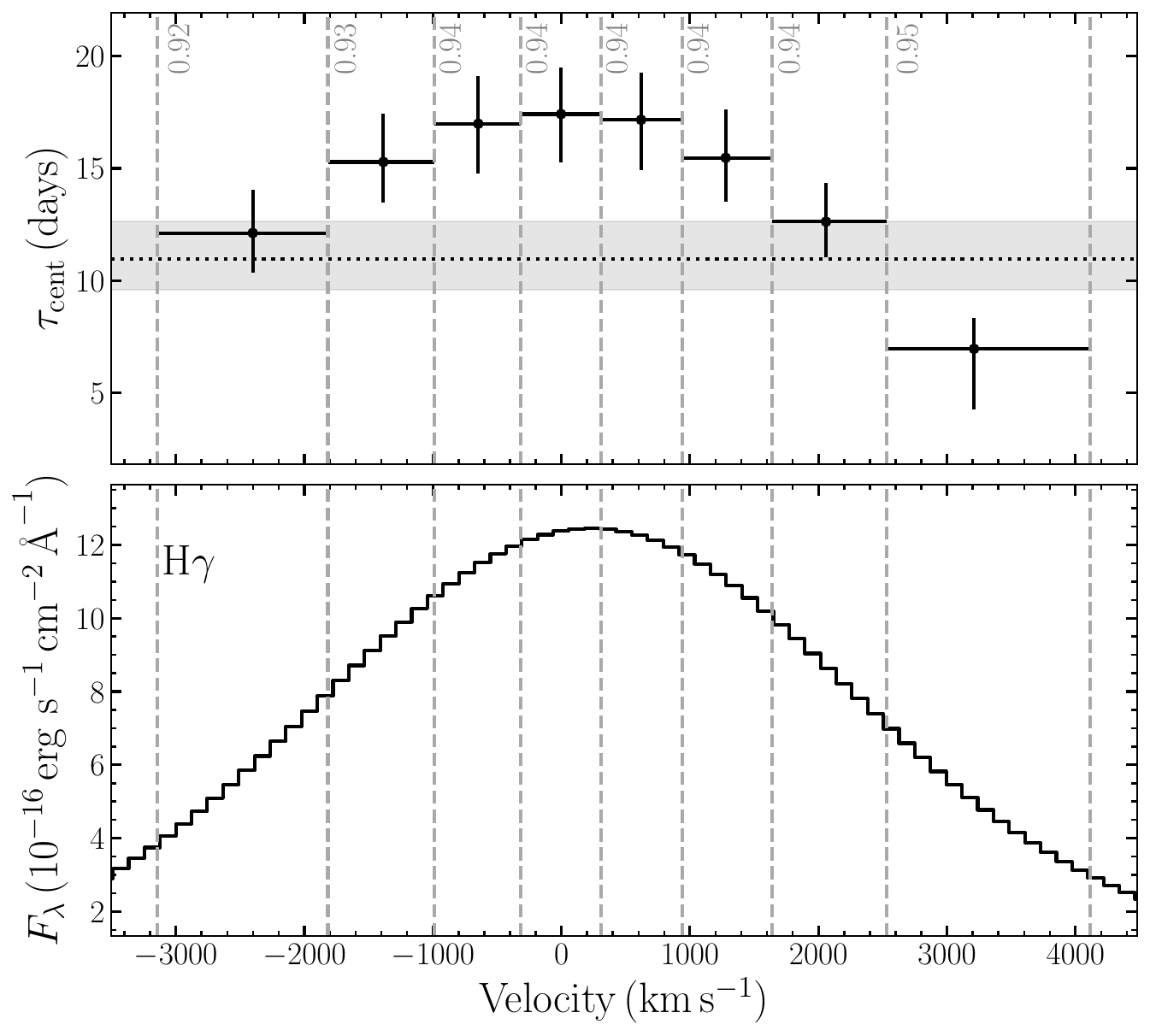}
\includegraphics[scale=0.26]{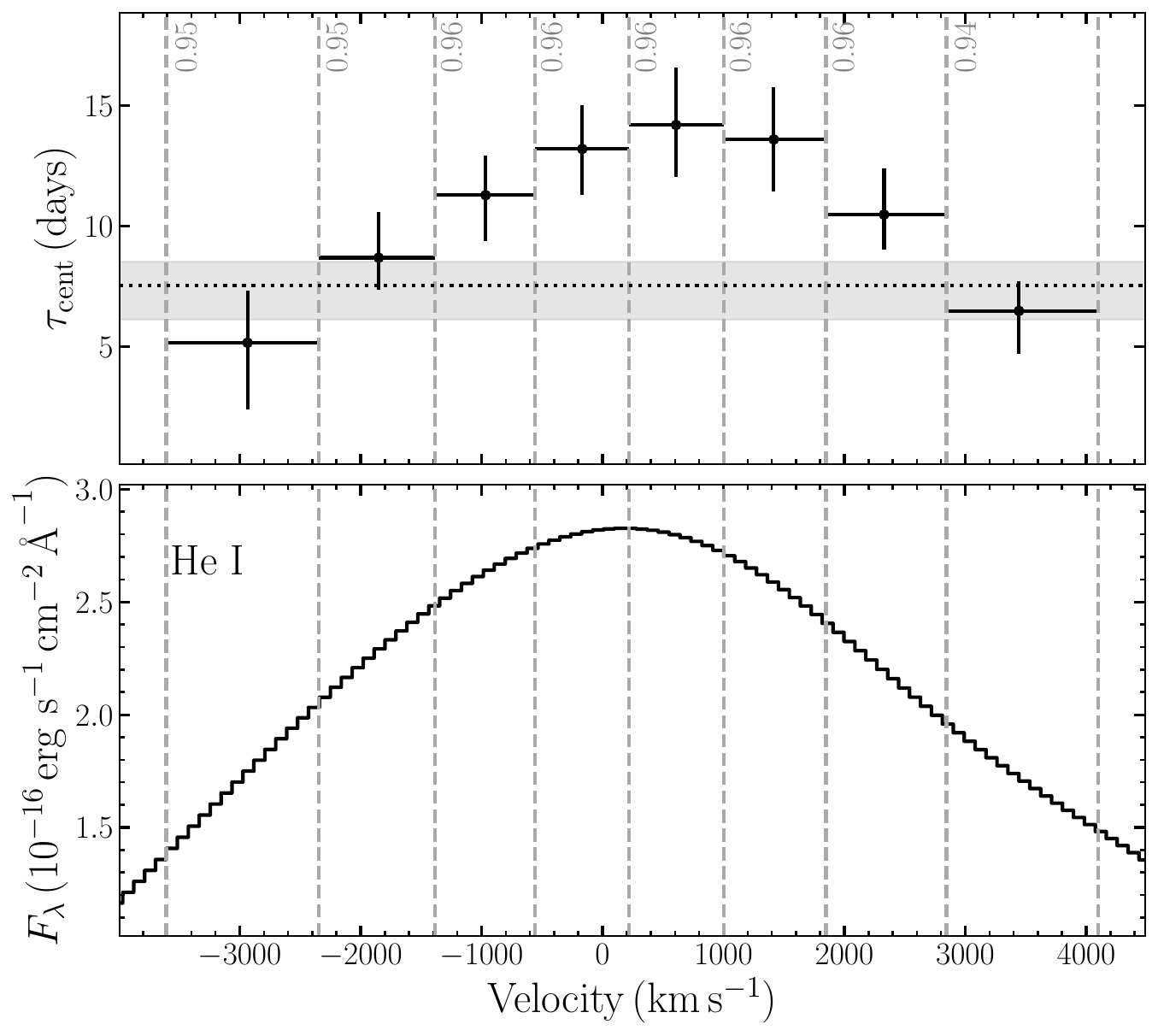}
\includegraphics[scale=0.26]{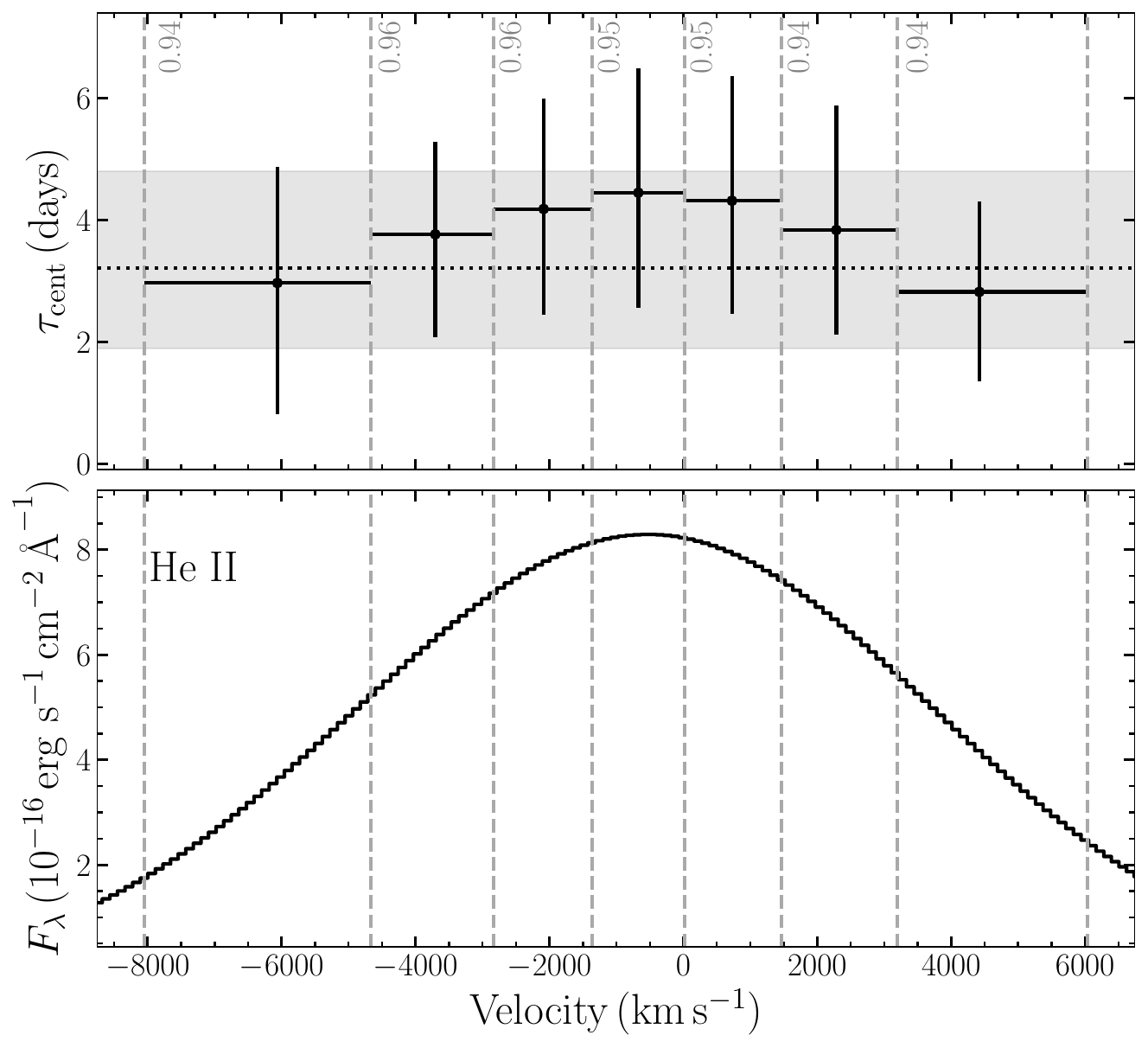}
\caption{Same as Figure~\ref{fig:vr1}, but for UGC~3374.}
\label{fig:vr2}
\end{figure*}

We calculated the centroid time lag ($\tau_{\rm cent}$) using points where the CCF value exceeded 0.8 times the peak value ($r_{\rm max}$), with $\tau_{\rm cent}$ serving as our measurement of the time lag. To estimate the uncertainties associated with the time lag, we utilized Monte Carlo simulations based on the ``random subset selection" and ``flux randomization" methods \citep{Peterson1998, Peterson2004}. These simulations were repeated 10,000 times to construct a cross-correlation centroid distribution (CCCD). The 15.87\% and 84.13\% percentiles of the CCCD were adopted as the lower and upper bounds of the time lag uncertainty, respectively. The final results of the CCF analysis, along with the autocorrelation function (ACF) of the photometry light curve, are shown in Figure~\ref{fig:speclc} and listed in Table~\ref{table:rm}, indicating an ionization-stratified BLR in both objects.

\begin{deluxetable*}{lcccccccccccc}[!ht]
 \tablecolumns{12}
\tablewidth{\textwidth}
\tabletypesize{\scriptsize}
\tablecaption{Time Lag, Line Width, Black hole Masses \label{table:rm}}	
\tablewidth{\textwidth}
\tablehead{& &   \multicolumn{2}{c}{CCF} & & \multicolumn{2}{c}{Mean} &&  \multicolumn{2}{c}{rms}  \\
\cline{3-4} \cline{6-7} \cline{9-10} 
\colhead{Object} & \colhead{Line} & \colhead{$r_{\rm max}$} & \colhead{$\tau_{\rm cent} $} &\colhead{}& \colhead{$\rm FWHM$} & \colhead{$\sigma_{\rm line}$} &&   \colhead{$\rm FWHM$} & \colhead{$\sigma_{\rm line}$} & \multicolumn{1}{@{}c@{}}{$M_{\rm VP}$} &\multicolumn{1}{@{}c@{}}{$M_{\bullet}$}\\ 
& &  & \multicolumn{1}{c}{(days)}  &&\multicolumn{2}{c}{($\kms$)} &&\multicolumn{2}{c}{($\kms$)} &  \multicolumn{2}{c}{($\times 10^7 M_{\odot}$)}
   }		
\startdata
KUG~1141+371& \ha & 0.88 &  $12.17_{-2.57}^{+0.69}$  && $5371 \pm 27$ & $2288 \pm 11$ && $4483 \pm 151$ & $2470 \pm 77$ & $1.45_{-0.32}^{+0.12}$ &$6.48_{-2.31}^{+1.89}$ \\
& \hb & 0.89 &  $7.79_{-3.65}^{+0.77}$  && $6387 \pm 53$ & $2731 \pm 22$ && $5848 \pm 291$ & $2965 \pm 80$ & $1.34_{-0.63}^{+0.15}$ &$5.98_{-3.27}^{+1.80}$ \\
& \hg & 0.79 &  $6.99_{-7.25}^{+3.95}$  && $6381 \pm 56$ & $2738 \pm 23$ && $6022 \pm 249$ & $2886 \pm 77$ & $1.14_{-1.18}^{+0.64}$ &$5.08_{-5.46}^{+3.21}$ \\
& \hei & 0.74 &  $0.21_{-4.95}^{+4.70}$  && $8679 \pm 210$ & $3922 \pm 57$ && $12656 \pm 614$ & $4793 \pm 117$ & $0.10_{-2.22}^{+2.11}$ &$0.43_{-9.92}^{+9.42}$ \\
& \heii & 0.64 &  $3.84_{-7.71}^{+4.67}$  && $9014 \pm 250$ & $4015 \pm 65$ && $9790 \pm 497$ & $4330 \pm 62$ & $1.41_{-2.82}^{+1.71}$ &$6.28_{-12.73}^{+7.83}$ \\
\hline
UGC~3374 & \ha & 0.96 &  $23.80_{-2.10}^{+2.05}$  && $4741 \pm 14$ & $2010 \pm 5$ && $3766 \pm 36$ & $1823 \pm 22$ & $1.54_{-0.14}^{+0.14}$ &$6.90_{-2.03}^{+2.03}$ \\
& \hb & 0.95 &  $9.57_{-3.29}^{+1.16}$  && $5270 \pm 26$ & $2247 \pm 11$ && $4853 \pm 142$ & $2488 \pm 48$ & $1.16_{-0.40}^{+0.15}$ &$5.17_{-2.30}^{+1.59}$ \\
& \hg & 0.94 &  $10.15_{-2.43}^{+1.73}$  && $5265 \pm 26$ & $2243 \pm 11$ && $4994 \pm 99$ & $2351 \pm 37$ & $1.10_{-0.26}^{+0.19}$ &$4.89_{-1.81}^{+1.61}$ \\
& \hei & 0.95 &  $4.76_{-3.49}^{+1.20}$  && $6292 \pm 62$ & $2727 \pm 30$ && $8682 \pm 548$ & $3577 \pm 94$ & $1.19_{-0.87}^{+0.31}$ &$5.31_{-4.18}^{+2.02}$ \\
& \heii & 0.97 &  $0.68_{-2.69}^{+1.93}$  && $9250 \pm 95$ & $3995 \pm 31$ && $10125 \pm 206$ & $4291 \pm 48$ & $0.24_{-0.97}^{+0.70}$ &$1.09_{-4.33}^{+3.12}$ \\
\enddata
\tablecomments{The time lags are measured in the rest frame, and line widths are corrected for instrumental broadening. $M_{\rm VP}$ represents the virial product, based on the $\sigma_{\rm line}$ measured from the rms spectrum and $\tau_{\rm{cent}}$. We adopted $f = 4.47\pm 1.25$ from \citet{Woo2015} to measure \mbh, incorporating the uncertainties associated with time lag, line width, and $f$.
}
\end{deluxetable*}

\subsection{Line Width and Black Hole Mass} \label{sec:4.2}
We determine the velocity of the BLR using the full-width at half maximum (FWHM) and line dispersion ($\sigma_{\rm line}$), both of which are measured from the mean and rms spectra. Following the approach outlined in \citet{Feng2024}, we generate the mean and rms spectra using the broad-line-only spectra obtained through spectral fitting, thereby minimizing contamination from narrow lines. To estimate the final values and uncertainties of FWHM and $\sigma_{\rm line}$, we employ a bootstrap technique. Specifically, $N$ spectra are randomly selected, with replacement, from the original set of $N$ spectra. Duplicate spectra are removed before generating new mean and rms spectra. This process is repeated 1000 times. The median values of these measurements were adopted as the velocity estimates, while the standard deviations were used to quantify the uncertainties.

To account for instrumental broadening, we compare the fitted width of \oiii\,$\lambda$5007 with that from a higher resolution spectrum. For KUG 1141+371, the intrinsic FWHM of \oiii\,$\lambda$5007 is calculated using SDSS data, with the instrumental broadening estimated to be approximately 1200 $\kms$. In the case of UGC~3374, the intrinsic FWHM of \oiii\,$\lambda$5007 was reported as 605 $\kms$ by \citet{Whittle1992}, and the instrumental broadening is similarly estimated to be around 1200 $\kms$. The final velocity measurements, corrected for instrumental broadening, are listed in Table \ref{table:rm}.

By combing time lags and line widths, we can estimate the \mbh\ using the following equation 
\begin{equation}
\mbh = f\frac{(c\tau)V^2}{G} \equiv fM_{\rm VP},
\end{equation}
where $G$ is the gravitational constant, $c$ is the speed of light, $\tau$ is the time lag, and $V$ is the line width. Here, $c\tau$ represents the mean size of the BLR. The virial factor, $f$, is influenced by the geometry, kinematics, and inclination of the BLR. This factor can be determined through dynamical modeling \citep{Pancoast2012, Li2018} for individual targets or calibrated using \mbh-$\sigma_{*}$ relation \citep{Onken2004, Woo2015} for RM samples. In this work, we used $\sigma_{\rm line}$ measured from rms spectrum and adopt $f = 4.47\pm 1.25$ from \citet{Woo2015} to estimate \mbh, incorporating the uncertainty of $f$, as illustrated in Table~\ref{table:rm}. For both targets, the \mbh\ values derived from different emission lines are generally consistent within their respective uncertainties, but a trend is observed where shorter time lags correspond to smaller black hole masses.

\subsection{Velocity-resolved Delays}
To investigate the overall geometry and kinematics of the BLR, we measured velocity-resolved lags for multiple emission lines. Using the rms spectrum described in Section \ref{sec:4.2}, we divided each emission line was into several velocity bins, with each bin designed to contain an equal amount of flux. Light curves corresponding to these velocity bins are extracted from the individual spectra, and time lags for each velocity bin are determined using the interpolated CCF method. Minor reverberation effects within individual bins are neglected in this analysis \citep{DeRosa2018}. The results are presented in Figure \ref{fig:vr1} for KUG 1141+371 and Figure \ref{fig:vr2} for UGC 3374.

KUG 1141+371 exhibits a symmetric velocity-resolved lag profile for the \ha\ emission line, which originates from the outer regions of the BLR. In contrast, the \hb, \hg, \hei\ and \heii\ lines, which trace the inner regions, display longer lags on the red side than on the blue side. UGC 3374, however, exhibits a different behavior. The velocity-resolved lag profiles of \hb, \hg, \hei\ and \heii\ lines are nearly symmetric, whereas the \ha\ line shows longer lags on the blue side than on the red side. These patterns indicate that the geometry and/or kinematics of the BLR vary with distance from the central black hole for both targets.

\section{Discussion} \label{sec:discuss}

\subsection{$R-L$ Relation}
One of the primary goals in RM studies is to establish an a reliable relationship between AGN luminosity and the size of the BLR. However, despite efforts over the past decade to expand the sample size and improve the quality of observational data, the scatter in the $R-L$ relation has increased. \citet{Du2018} reported that for super-Eddington accreting sources, the size of the \hb\ emitting region ($R_{\rm{H}\beta}$) is shorter than that expected from the $R-L$ relation. Furthermore, our recent findings suggest that the time delay between the UV and optical continuum may have a similar effect \citep{Feng2024}. These results imply that the scatter may be linked to the intrinsic physical properties of AGNs.

\begin{figure}[!ht]
\centering 
\includegraphics[scale=0.6]{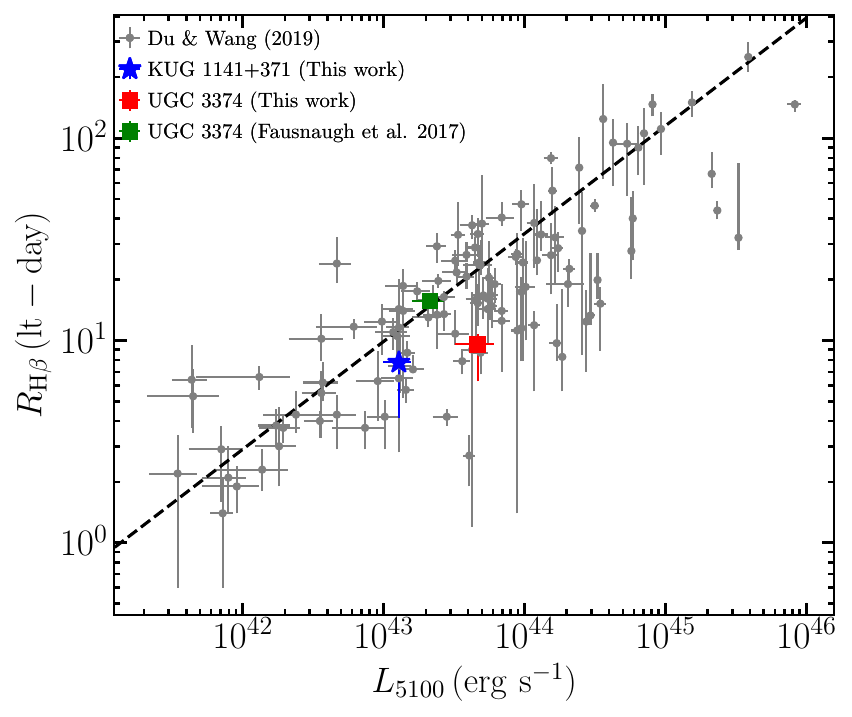}
\caption{$R_{\rm H\beta}$ - $L_{\rm 5100}$ relation. The gray points correspond to RM results from \citet{Du2019}. The blue star represents KUG~1141+371, while the red and green squares represent \hb\ results of UGC~3374 obtained from this work and \citet{Fausnaugh2017}, respectively. The black dashed line represents $R_{\rm H\beta}$ - $L_{\rm 5100}$ relation as derived from \citet{Bentz2013}.
}
\label{fig:rl}
\end{figure}

This work presents the first RM observations for KUG 1141+37, while UGC 3374 was previously monitored in 2014. To examine the $R-L$ relation for these two sources, we plot their $R_{\rm{H}\beta}$ against $L_{\rm 5100}$ (the monochromatic luminosity at 5100 \AA) in Figure \ref{fig:rl}. For comparison, we also include the RM sample from \citet{Du2019}. $L_{\rm 5100}$ is calculated using the mean flux of the fitted continuum at 5100 \AA\ (see Section \ref{sec:3.1}), yielding log $L_{\rm 5100}$ values of $43.11 \pm 0.10$ and $43.67 \pm 0.14$ ${\rm erg \, s}^{-1}$ for KUG 1141+371 and UGC 3374, respectively. Combined with their black hole masses, the corresponding dimensionless accretion rates\footnote{The dimensionless accretion rate ($\dotm$) is defined as $\dot{\mbh} c^2 / L_{\rm Edd}$, where $\dot{\mbh}$ is the mass accretion rate and $L_{\rm Edd}$ is the Eddington luminosity, following the convention in \citet{Du2018}.} are estimated to be $0.04$ and $0.38$, indicating that both AGNs fall within the low-accretion regime. 

Low-accretion rate AGNs are generally found to follow the $R-L$ relation established by \citet{Bentz2013}:
\begin{equation}
{\rm log}\, \frac{R_{{\rm H}\beta}}{\rm 1\, \mathrm{lt\text{-}day}} = 1.527^{+0.031}_{-0.031} + 0.533^{+0.035}_{-0.033}\,{\rm log}\, L_{44},
\end{equation}
where $L_{44}$ is the monochromatic luminosity $L_{\rm 5100}$ in units of $10^{44} {\rm erg \, s}^{-1}$. Based on this relation, the expected time delays for KUG 1141+371 and UGC 3374 are 11.37 days and 22.60 days, respectively, which are 1.46 and 2.36 times larger than the observed values.

\citet{Du2019} proposed that the flux ratio of \feii\ to \hb\ (\Rfe = $F_{\rm Fe\ II }$ / $F_{\rm{H}\beta }$) can mitigate the shortened lag issue for high-accretion rate AGNs. They introduced a modified $R-L$ relation, which incorporates ${\cal R}_{\rm Fe}$ as an additional parameter:
\begin{equation}
\begin{split}
{\rm log}\, \frac{R_{{\rm H}\beta}}{\rm 1\, \mathrm{lt\text{-}day}} = &1.65^{+0.06}_{-0.06} + 0.45^{+0.03}_{-0.03}\,{\rm log}\, L_{44}\\
                                           &-0.35^{+0.08}_{-0.08}\,{\cal R}_{\rm Fe}.                                        
\end{split}
\label{eq5}
\end{equation} 
Using the fitted results, we obtain the total flux of \feii\ within the 4434 \AA\ to 4684 \AA\ range and infer median ${\cal R}_{\rm Fe}$ values of 0.31 and 0.34 for KUG 1141+371 and UGC 3374, respectively. Substituting these values into Equation(\ref{eq5}), the predicted lags are 13.78 days and 24.09 days, which further deviate our observations. This discrepancy may be attributed to the fact that the modified $R-L$ relation from \citet{Du2019} may be more suitable for high-accretion rate AGNs, or it could imply that the BLR physical properties of these two targets are indeed unusual.

\begin{figure}[!ht]
\centering 
\includegraphics[scale=0.43]{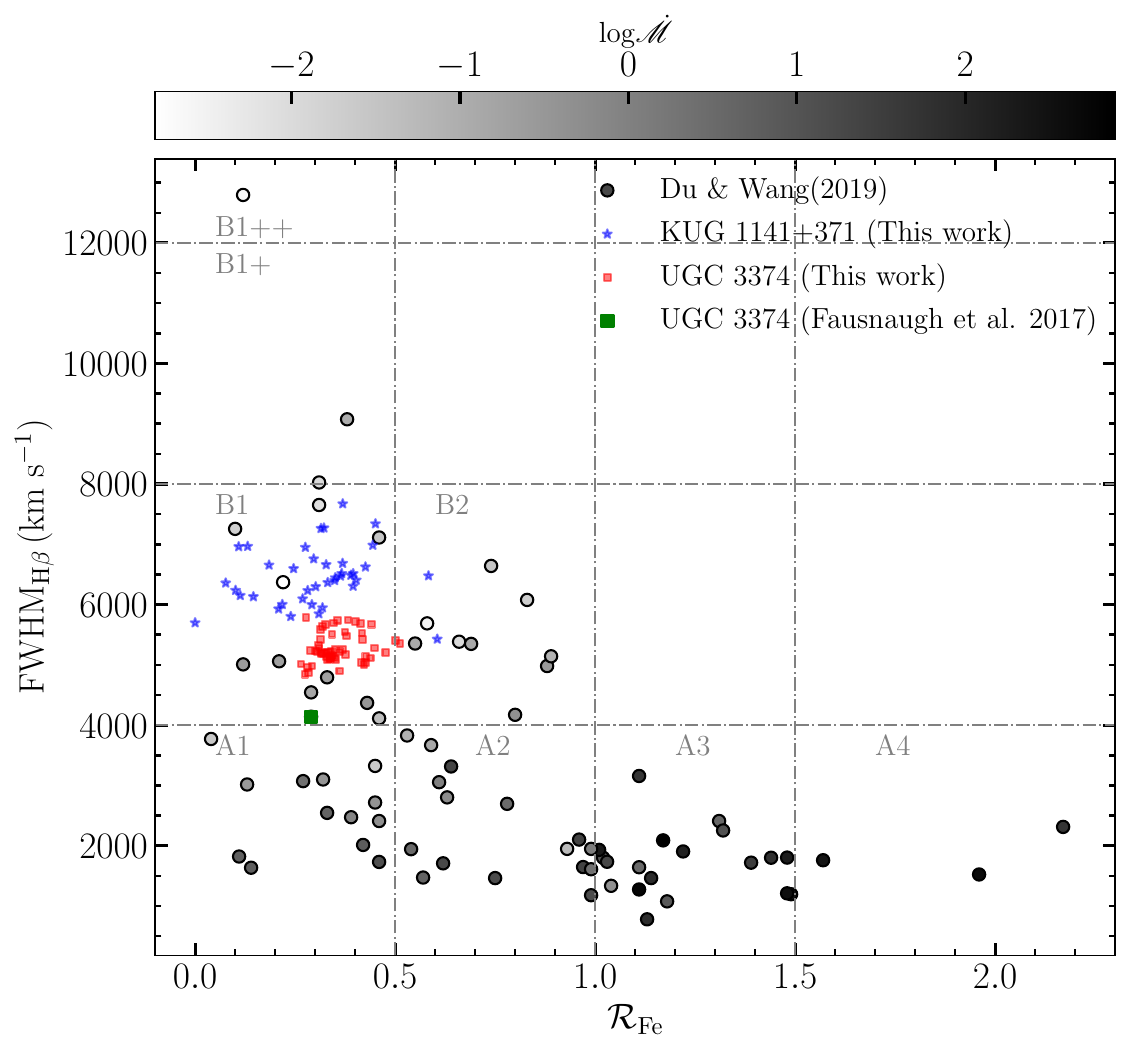}
\caption{The Eigenvector 1 plane of AGNs. The dots represent the sample from \citet{Du2019}, color-coded by log $\dotm$. The green square marks the result of UGC 3374 from \citet{Fausnaugh2017}. The blue stars and red squares display the results derived from the individual spectra of KUG 1141+371 and UGC 3374, respectively. Both targets are located within the Population B1 region.
}
\label{fig:rfe-fwhm}
\end{figure}

To investigate whether KUG 1141+371 and UGC 3374 are peculiar AGNs, we compare their positions on the Eigenvector 1 plane \citep{Boroson1992, Marziani2018, Panda2020} with those of the RM sample. As shown in Figure \ref{fig:rfe-fwhm}, both sources are primarily located in the Population B1 region during our monitoring period, which is typical for AGNs with low accretion rates \citep{Bon2018, Panda2024}. The shorter-than-expected lags observed in KUG 1141+371 and UGC 3374 suggest that additional factors beyond the accretion rate may influence the $R-L$ relation.

\subsection{Stratified Geometry and Kinematics in BLR}
Both KUG 1141+371 and UGC 3374 exhibit clear radial ionization stratification in their BLRs, consistent with photoionization simulations and other RM results \citep{Korista2004, Fausnaugh2017, Feng2021b, Feng2024}. This indicates that each emission line predominantly originates from a specific region within the BLR. By measuring velocity-resolved lags for different emission lines in these two AGNs, we can explore the geometric and kinematic properties of different regions (radii) within the BLR.

\subsubsection{Empirical Recipes for Various Velocity-resolved Lags}
The shape of velocity-resolved lags may be related to the geometry, structure, and kinematics of the BLR. RM studies generally attribute these lag profiles to different kinematic regimes, assuming a symmetric BLR geometry and neglecting internal substructures \citep{Welsh1991, Bentz2010}. For instance, Keplerian rotation or virialized motion will result in symmetric velocity-resolved lags, where the lags are shortest at the highest velocities (both redshifted and blueshifted) and longest at lower velocities near the line center. In contrast, the asymmetric velocity-resolved lags can indicate non-virialized kinematics: longer lags on the blueshifted side suggest inflowing gas, while longer lags on the redshifted side often imply outflowing gas. Within this framework, the distinct velocity-resolved lag profiles observed for different BELs in KUG 1141+371 and UGC 3374 suggest that their BLRs exhibit different kinematics in the inner and outer regions (as shown in Figure \ref{fig:vr1} and \ref{fig:vr2}).

In KUG 1141+371, the velocity-resolved lags of the inner BLR, traced by \hb, \hg, \hei, and \heii, show longer lags on the red side of the emission lines, indicating an outflow in these inner regions. The redshifted asymmetry implies that gas is being driven outward from the vicinity of the central engine, possibly due to radiation pressure or other non-gravitational forces. The outer BLR, as traced by \ha, exhibits symmetric velocity-resolved lags, which are consistent with virialized motion. This suggests that the gas at larger radii is predominantly gravitationally bound and follows Keplerian orbits around the central black hole.

For UGC 3374, a different kinematic pattern emerges. The velocity-resolved lags of the outer BLR, traced by \ha, show longer lags on the blue side of the emission lines, suggesting the presence of inflowing gas in the outer regions. This blueshifted asymmetry indicates that gas is moving inward from the outskirts of the BLR toward the central regions. Meanwhile, the other emission lines, which trace the inner BLR, exhibit symmetric velocity-resolved lags, indicating that the gas in this region is primarily in virialized motion.

However, the observed asymmetries in the velocity-resolved ionization properties challenge the assumption of simple kinematic mechanisms, which require an asymmetric geometry within the BLR \citep{Li2024}. In this scenario, the asymmetric velocity-resolved lag profiles could naturally arise if the redshifted and blueshifted gas are located at varying distances from the central ionizing source. Additionally, single kinematic variations are difficult to interpret the variability of velocity-resolved lags. As the radiation pressure increases, the BLR exhibits a transition from outflow-dominated to virialized motion, with the outer regions showing a more rapid variation than the inner regions \citep{Feng2024}. Simulations by \citet{Du2023} have demonstrated that anomalous distribution of gas within the BLR can indeed account for complex velocity-delay maps. Therefore, it is plausible that the inner and outer regions of the BLRs in KUG 1141+371 and UGC 3374 may not only exhibit distinct kinematics but also present different geometries or structures.

\subsubsection{Interpretations of the Stratified BLR Geometry and Kinematics}
The distinct geometry and kinematics observed in the inner and outer regions of the BLR may be attributed to mechanisms related to the origin of the BLR. A plausible explanation is that during the formation and evolution of the BLR, the gas can be accelerated or decelerated, resulting in complex structures. In this section, we focus on the mechanisms associated with the origin of the BLR, assuming that the observed velocity-resolved lags primarily reflect the gas kinematics.

Two major origins of the BLR gas are often considered: the accretion disk or the dusty torus. If the BLR gas originates from the accretion disk, it is thought that the gas reach to BLR via hydromagnetically or/and radiatively accelerated outflow winds \citep{Emmering1992, Murray1995, Proga2004}. Alternatively, if the BLR gas originates from the dusty torus, the dust clouds may be tidally disrupted by the central SMBH or undergo failed radiatively accelerated dusty outflows \citep{Czerny2011, Wang2017, Naddaf2021}, generating new BLR gas. Below, we discuss the possible kinematic structures produced by each mechanism:

\begin{enumerate}

\item Radiation pressure-driven models: If both radiation pressure and gravitational force decrease with distance from the central black hole as \(R^{-2}\), and the outflow is line-driven, the BLR gas will be continuously accelerated from the inner to outer regions, resulting in progressively stronger outflows. However, if gas absorption and scattering cause radiation pressure to decrease faster than \(R^{-2}\), outflows may only exist in the inner BLR, while the outer BLR would be dominated by gravitational forces, leading to virialized motion or even inflows. This scenario is consistent with the observations of KUG~1141+371.

\item Magnetically driven models: Magnetic pressure can effectively drive outflows within the accretion disk, but its influence weakens in the BLR. As a result, the gas is gradually decelerated by gravity, leading to virialized motion. This model can also explain the phenomena observed in KUG 1141+371, where outflows dominate in the inner BLR, while virialized gas is present in the outer regions.

\item Tidal disruption of dust clouds: When dust clouds in the torus are tidally disrupted, they spiral inward, potentially resulting in accelerated inflows. These inflows are typically more significant in the inner BLR, as the gas moves closer to the black hole. On the other hand, during this inward migration, friction or radiation pressure may cause the gas to gradually transition to circularized motion or even outflows, consistent with the observations of UGC 3374.

\begin{figure*}[!ht]
\centering 
\includegraphics[scale=0.4]{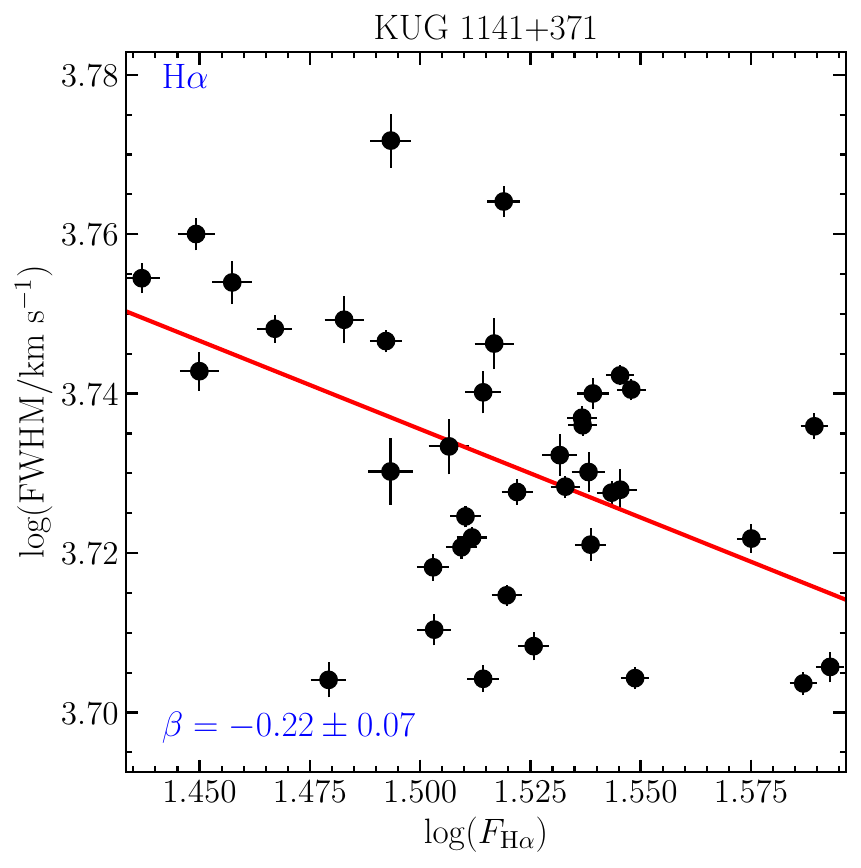}
\includegraphics[scale=0.4]{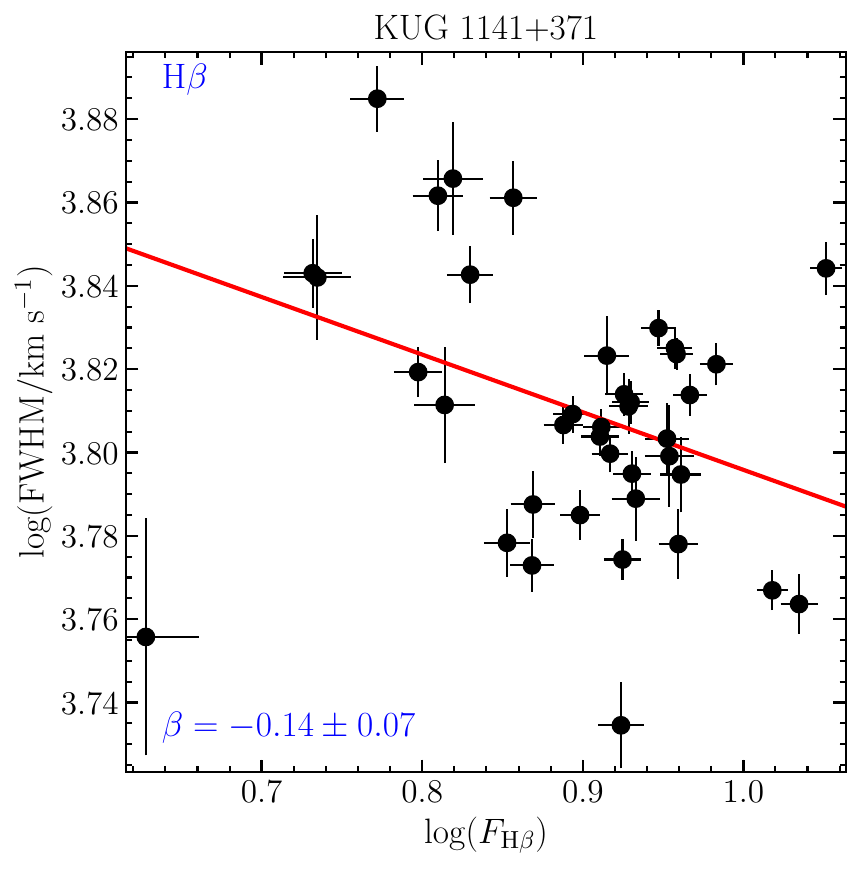}
\includegraphics[scale=0.4]{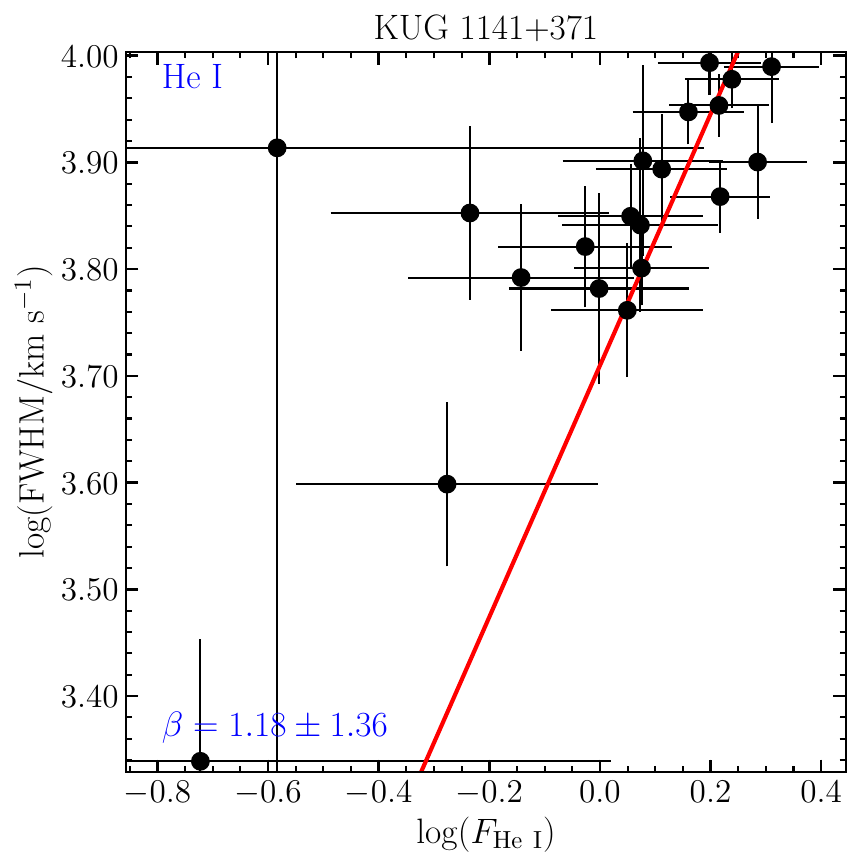}
\includegraphics[scale=0.4]{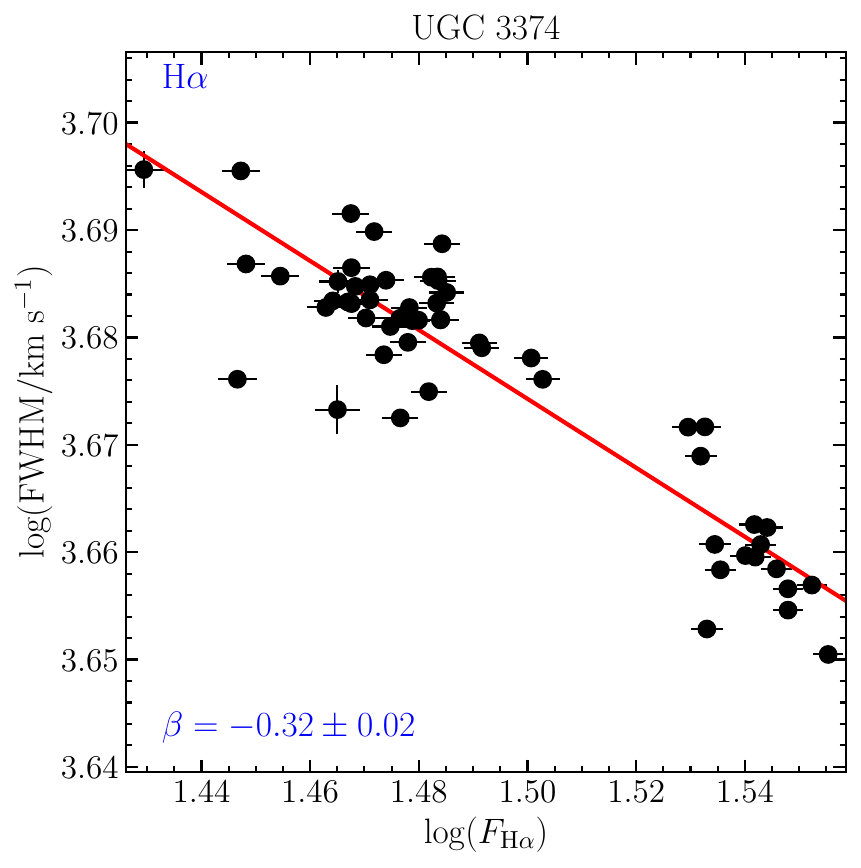}
\includegraphics[scale=0.4]{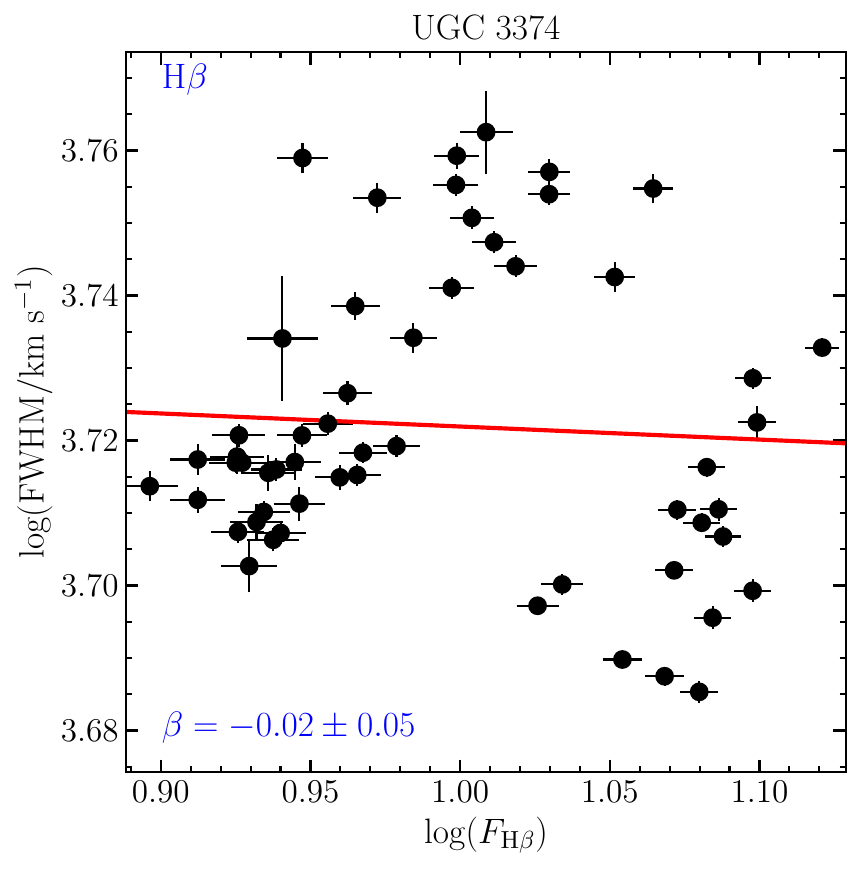}
\includegraphics[scale=0.4]{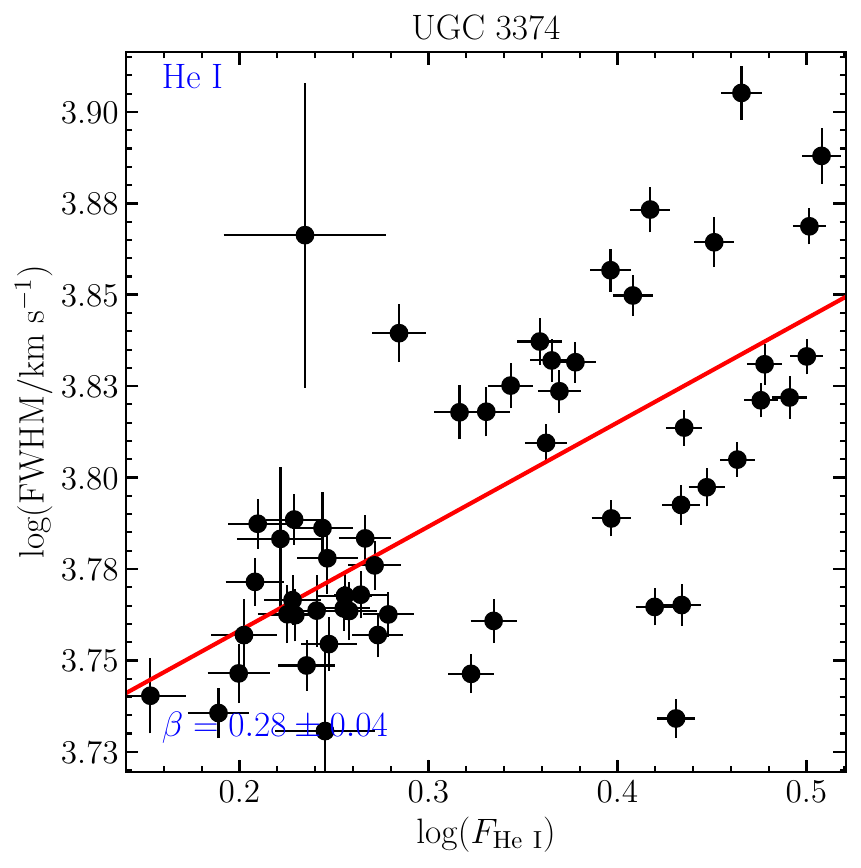}
\caption{The relation between the emission line flux and its FWHM. Each panel presents the results for \ha, \hb, and \hei\ for KUG 1141+371 and UGC 3374, respectively. Black dots indicate the best-fit results derived from individual spectra, while the red line represents the best linear fit obtained using linmix \citep{Kelly2007}.   
}
\label{fig:fwhmflux}
\end{figure*}

\item Failed radiatively accelerated dusty outflows: Some shielded dust grains are radiatively accelerated, forming outflows. As the dust reaches a sublimation location, the grains are destroyed due to the increased illuminated surface area, and radiation pressure diminishes. Consequently, the gas may experience inflows or virialized motion under the influence of gravity, leading to a producing of different kinematic signatures within the BLR.

\end{enumerate}

It is worth noting that the observed asymmetries in velocity-resolved lags may simply result from complex geometric structures within the BLR, such as elliptical disks or spiral arms, rather than indicating non-virialized kinematics \citep{Li2024}. This interpretation is further supported by RM studies of NGC 4151, where asymmetric velocity-resolved lags have been detected across multiple emission lines. Nevertheless, over several years of monitoring, the velocity-lag relationships in different lines have remained consistent with the virial relation \citep{Li2022, Chen2023, Feng2024}. Therefore, future long-term, multiline RM observations, combined with simultaneous multiwavelength continuum monitoring, will be crucial for disentangling the effects of geometry and kinematics within the BLR.

\subsection{Anti-breathing Effect}
Interestingly, during our observations, the luminosity of UGC 3374 increased by more than a factor of two compared to 2014, but its \hb\ lag ($\tau_{\rm H\beta}$) shortened to $\sim$62\% of the previous measurements. This so-called ``anti-breathing" phenomenon is generally detected in the \civ\ emission line and is interpreted as the possible presence of a non-varying component in the emission line \citep{Wang2020}. Another plausible explanation is that the UV-optical lag ($\tau_{\rm uv-opt}$) may change significantly with luminosity, and when the AGN enter into a bright state, this lag would be comparable to the \hb\ lag \citep{Zhou2024}. Since the gas in the BLR is photoionized by UV photons, the true $R_{\rm{H}\beta}$ should be ($\tau_{\rm uv-opt}$ + $\tau_{\rm H\beta}$) $\times c$. Therefore, directly using $\tau_{\rm H\beta}$ to infer $R_{\rm{H}\beta}$ may result in an underestimation, as demonstrated by recent multiline RM observations of NGC 4151 \citep{Feng2024}. In the faint state, the UV-optical lag of UGC 3374 is about 1 day \citep[see Figure 5 of][]{Fausnaugh2018}, which has a negligible effect on the BLR size measurement. Unfortunately, during our observations, simultaneous UV continuum light curves are unavailable, limiting our ability to give a definitive conclusion.

If the BLR is virialized ($V \propto R^{-0.5}$) and the emission-line flux scales with the size of the emitting region as $F \propto R^{0.5}$, the normal breathing behavior would predict $V \propto F^{-0.25}$. This implies that if the anti-breathing originates from intrinsic BLR properties, we would expect a positive correlation between $V$ and $F$. Conversely, if the anti-breathing is caused by the UV-optical lag, the correlation between $V$ and $F$ would remain negative, as the true physical size of the BLR does not decrease. During our observations, the \hb velocity dispersion measured from the rms spectrum is $2488 \pm 48$ km s$^{-1}$, which is indeed larger than the value of $1466^{+102}_{-174}$ km s$^{-1}$ reported by \citet{Fausnaugh2017}. This suggests that the observed anti-breathing phenomenon may be linked to intrinsic properties of the BLR.

To further investigate the presence of an anti-breathing effect on short timescales (i.e., during our observing campaign), we examined the relationship between $\sigma_{\rm line}$ and $F$ for \ha, \hb, and \hei. We excluded \hg\ and \heii\ from this analysis because the line profile of the former is tied to that of \hb, while the latter suffers from a lower S/N. As shown in Figure \ref{fig:fwhmflux}, \ha\ in both AGNs follows the breathing behavior expected from virialized motion, while \hei\ exhibits a clear anti-breathing. The behavior of \hb\ falls between these two extremes. This result suggests significant differences in the physical properties of the inner and outer BLR regions, which may arise from variations in geometry, kinematics, or ionization conditions. The shortened time delay detected in UGC 3374 may reflect the fact that \hb\ resides in a transition region influenced by both breathing and anti-breathing behaviors.

\subsection{Implications for Reverberation Mapping Studies}
Recent RM studies have reported rapid changes in velocity-resolved lag profiles, and interpreted as indicators of evolving BLR geometry and kinematics \citep{Hu2020, Lu2022, Yao2024}. However, the variability timescales are shorter than the expected dynamical timescales, and the observed luminosity-dependent kinematic changes are inconsistent with the outflow behavior predicted by radiation pressure models. The distinct stratification in geometry and kinematics observed in the BLRs of KUG1141+371 and UGC3374 offers a compelling explanation for these rapid changes in velocity-resolved lags.

Given that emission from a single line would only represent a narrow ionization region within the BLR, and that the location of the line-emitting region can shift with luminosity variations, the observed geometry and kinematics may appear to change even if the overall BLR gas remain stable. Conversely, if the geometry and kinematics are identical throughout the BLR, one would expect to observe long-term stability in the velocity-resolved lags. Indeed, this expectation has been observed in some targets. For example, NGC 3516 exhibits consistent velocity-resolved lag profiles for \ha\ and \hb, with nearly identical results obtained in 2007 and 2019 \citep{Denney2009, Feng2021a}, although we cannot determine whether its BLR has undergone changes during this period. A more convincing example is Arp 151, which shows comparable velocity-resolved lag profiles for \ha, \hb, and \hg\ that remained constant across three separate observations conducted between 2008 and 2015 \citep{Bentz2010, Pancoast2018}.

Moreover, the diverse geometry and kinematics within the BLR may impact measurements of black hole mass. This implies that using different emission lines, or the same emission line at different luminosities, could yield discrepant black hole mass estimates.  Several studies have shown considerable scatter in black hole mass derived from RM observations conducted at different periods, as well as results obtained from different emission lines. While the UV-optical lags may contribute to these discrepancies \citep{Zhou2024}, the distinct geometry and kinematics in the inner and outer BLR regions could also play a role. Further investigations are necessary to validate these effects after correcting for the UV-optical lags.

\section{Conclusion} \label{sec:conclusion}
We report the multiline RM results of two variable AGNs, KUG 1141+371 and UGC 3374. This is the first RM observation for KUG 1141+371 and the first velocity-resolved time delay analysis for UGC 3374. Our analysis reveals a clear radial stratification in the ionization structure of the BLRs, consistent with photoionization model predictions. Velocity-resolved RM measurements indicate distinct geometry and kinematics between the inner and outer regions of the BLRs in both AGNs. Under the assumption that the velocity-resolved lags reflect the kinematics of BLR, KUG 1141+371 exhibits signatures of outflow in the inner BLR and virialized motion in the outer region, while UGC 3374 shows evidence of inflow in the outer BLR and virialized motion in the inner region. Additionally, during our observations, we detected a transition from breathing to anti-breathing behavior between the outer and inner BLR regions in both AGNs. This transition further highlights the significant differences in the physical properties of the inner and outer BLR regions.

We interpret these findings in the context of various BLR formation scenarios, discussing the potential roles of radiation pressure, magnetically driven winds, and the disruption of dusty gas clouds in shaping the BLR structure and kinematics. The observed radial stratification in geometry and kinematics may provide a natural explanation for the rapid changes in velocity-resolved RM signatures reported in recent studies, as well as the scatter in black hole mass estimates derived from different emission lines or at different luminosity states.

Our results highlight the importance of long-term, multiline RM campaigns in probing the complex structure and evolution of AGN BLRs. Future studies incorporating simultaneous multiwavelength continuum monitoring will be crucial for disentangling the effects of geometry and kinematics within the BLR and improving the accuracy of black hole mass measurements. As RM sample sizes continue to grow, a clearer picture of the BLR origin and its connection to the accretion disk and dusty torus will emerge, advancing our understanding of the physical processes governing the gas dynamics in the vicinity of SMBH.
\vspace{5mm}

We thank the anonymous referee for their constructive comments, which have improved the quality of this manuscript. We also thank Yubin Li for the discussion. This work is supported by National Key R\&D Program of China (No. 2021YFA1600404), the National Natural Science Foundation of China (grants No. 12303022, 12203096, 12373018, 11991051, 12322303, 12203041, and 12073068), Yunnan Fundamental Research Projects (grants No. 202301AT070358 and 202301AT070339), Yunnan Postdoctoral Research Foundation Funding Project, Special Research Assistant Funding Project of Chinese Academy of Sciences, the Natural Science Foundation of Fujian Province of China (No. 2022J06002), the Youth Innovation Promotion Association of Chinese Academy of Sciences (2022058), and the Young Talent Project of Yunnan Province, and the science research grants from the China Manned Space Project with No. CMS-CSST-2021-A06. We acknowledge the support of the staff of the Lijiang 2.4 m telescope. Funding for the telescope has been provided by Chinese Academy of Sciences and the People’s Government of Yunnan Province. Based on observations obtained with the Samuel Oschin Telescope 48-inch and the 60-inch Telescope at the Palomar Observatory as part of the Zwicky Transient Facility project. ZTF is supported by the National Science Foundation under Grants No. AST-1440341 and AST-2034437 and a collaboration including current partners Caltech, IPAC, the Oskar Klein Center at Stockholm University, the University of Maryland, University of California, Berkeley , the University of Wisconsin at Milwaukee, University of Warwick, Ruhr University, Cornell University, Northwestern University and Drexel University. Operations are conducted by COO, IPAC, and UW. This research has made use of the NASA/IPAC Extragalactic Database (NED), which is operated by the Jet Propulsion Laboratory, California Institute of Technology, under contract with NASA. Supported by the National Science Foundation under Grants No. AST-1440341 and AST-2034437 and a collaboration including current partners Caltech, IPAC, the Oskar Klein Center at Stockholm University, the University of Maryland, University of California, Berkeley , the University of Wisconsin at Milwaukee, University of Warwick, Ruhr University, Cornell University, Northwestern University and Drexel University. Operations are conducted by COO, IPAC, and UW.

\vspace{5mm}

\facility{YAO:2.4m}
\software{PyRAF \citep{Pyraf2012}, DASpec \citep{Du2024}, PyCALI \citep{Pycali2024}, linmix \citep{Kelly2007}.}

\appendix
\section{Evaluation of Spectral Fitting Results and Uncertainty Estimates Using Bootstrap Resampling} \label{sec:appecdix}
To evaluate the robustness of our spectral fitting results and associated uncertainty estimates, we adopted a Monte Carlo simulation approach similar to that used for measuring time lags, incorporating random subset selection and flux randomization. For each spectrum, the simulation procedure was repeated 50 times to generate a distribution of the fitting results for each parameter. The mean value of the resulting distribution was taken as the final measurement, while the standard deviation of the distribution was used to estimate the uncertainty.

\begin{figure*}[!ht]
\centering 
\includegraphics[scale=0.345]{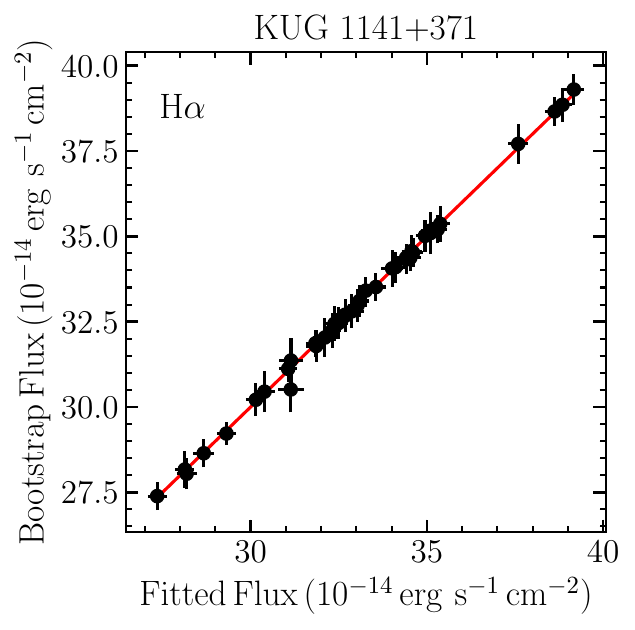}
\includegraphics[scale=0.345]{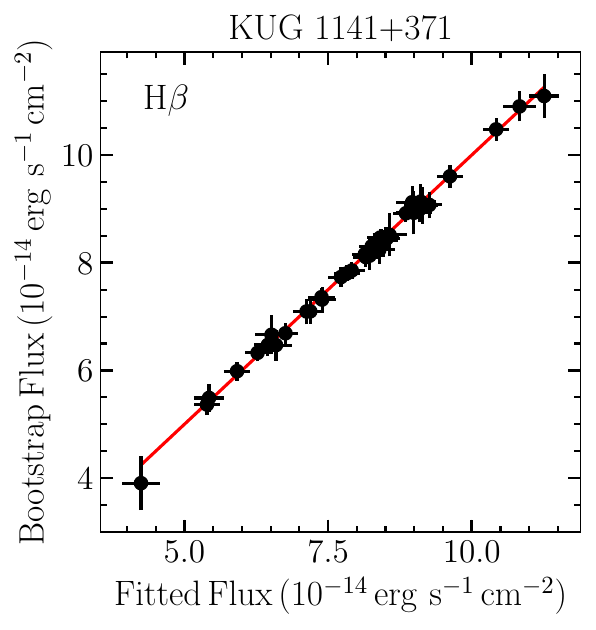}
\includegraphics[scale=0.345]{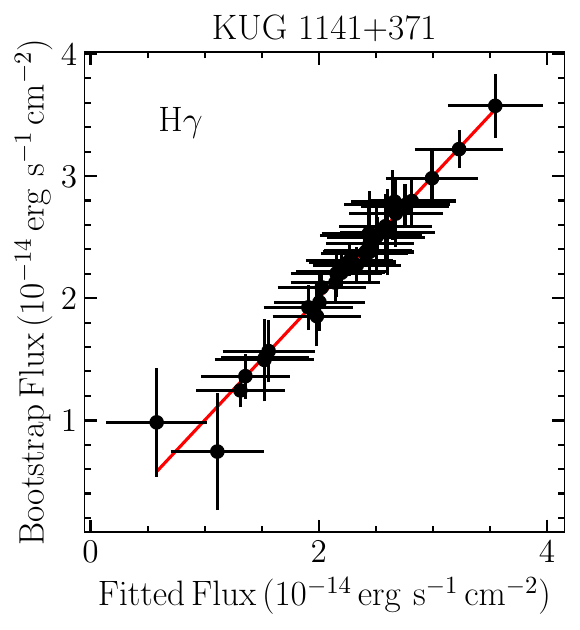}
\includegraphics[scale=0.345]{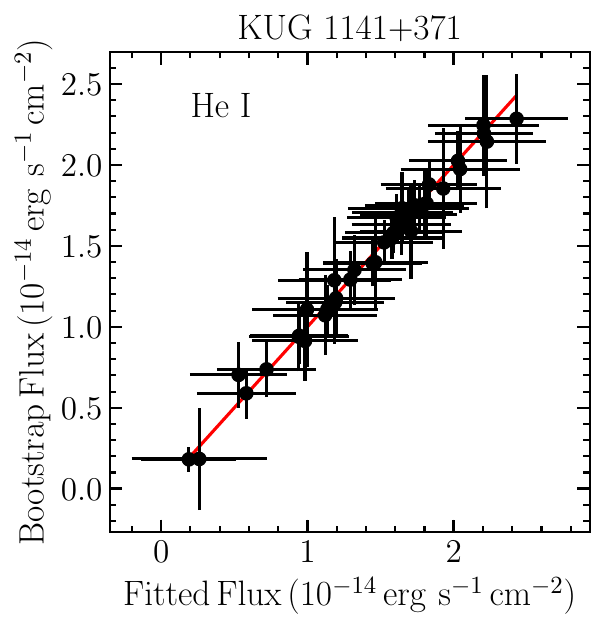}
\includegraphics[scale=0.345]{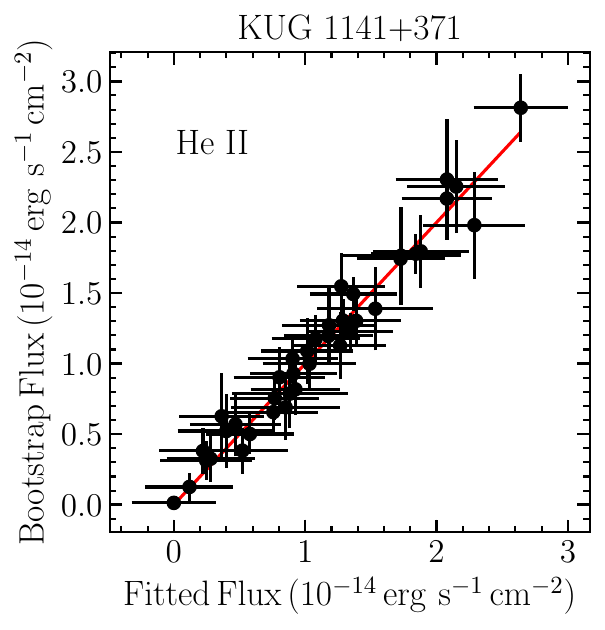}
\includegraphics[scale=0.345]{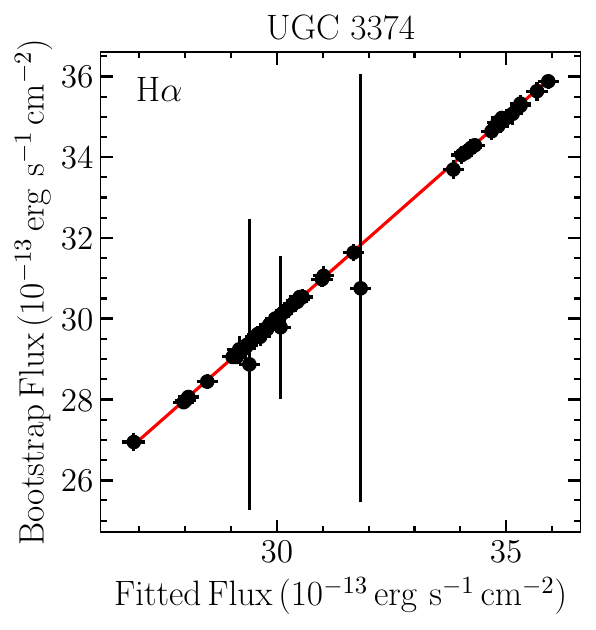}
\includegraphics[scale=0.345]{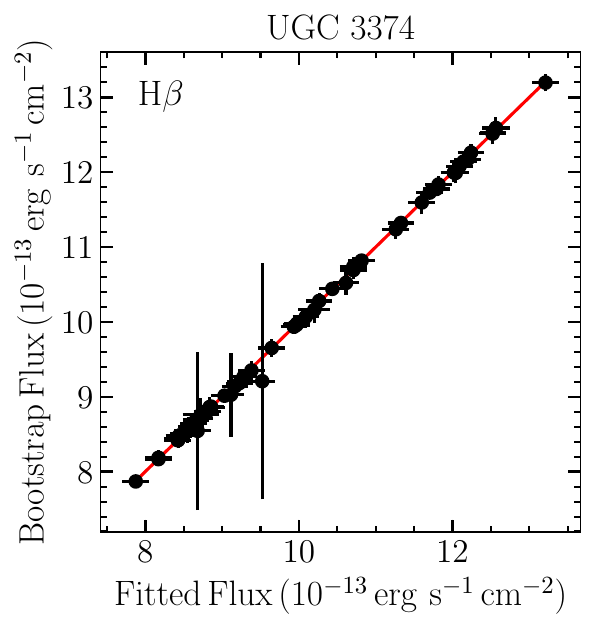}
\includegraphics[scale=0.345]{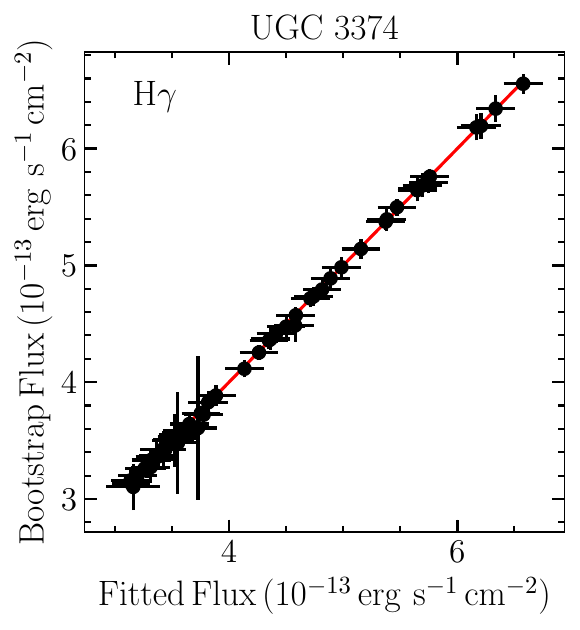}
\includegraphics[scale=0.345]{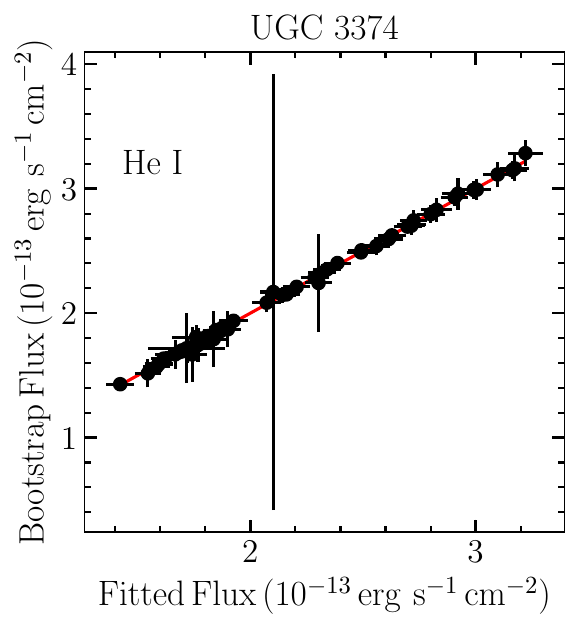}
\includegraphics[scale=0.345]{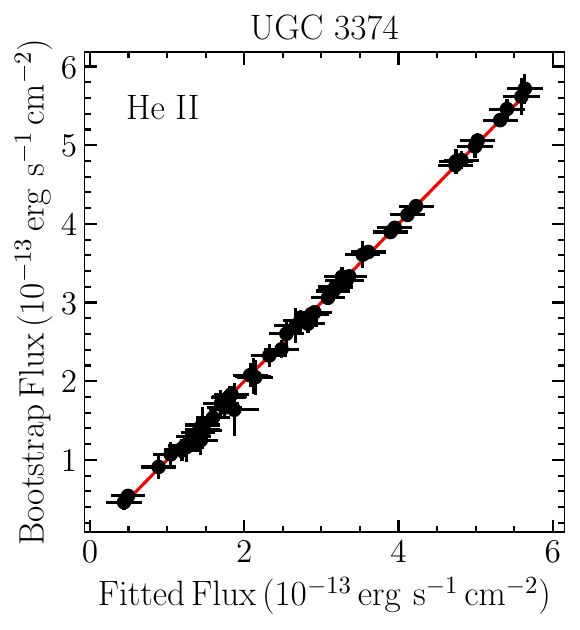}
\caption{Comparison of flux measurements obtained from individual spectra fitting and bootstrap resampling. ``Fitted Flux" refers to the flux determined from the original spectral fitting, while ``Bootstrap Flux" represents the flux estimated through the bootstrap resampling method. The red line represents the 1:1 relation, indicating perfect agreement between the two measurements.
}
\label{fig:comparisonflux}
\end{figure*}

\begin{figure*}[!ht]
\centering 
\includegraphics[scale=0.32]{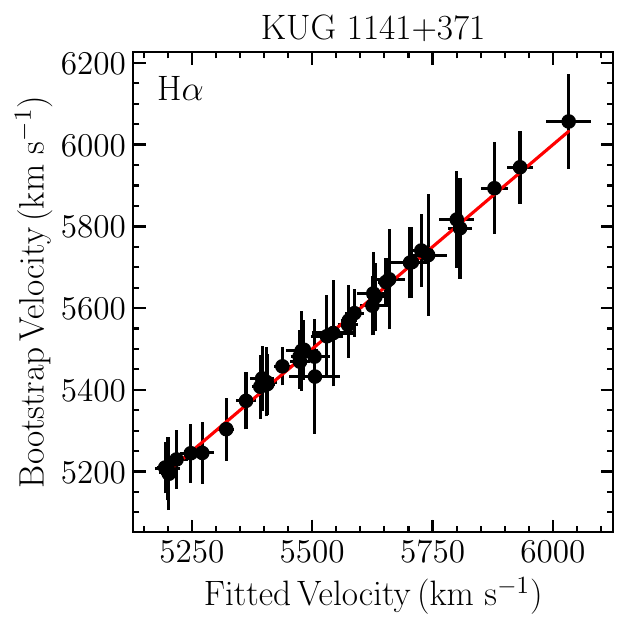}
\includegraphics[scale=0.32]{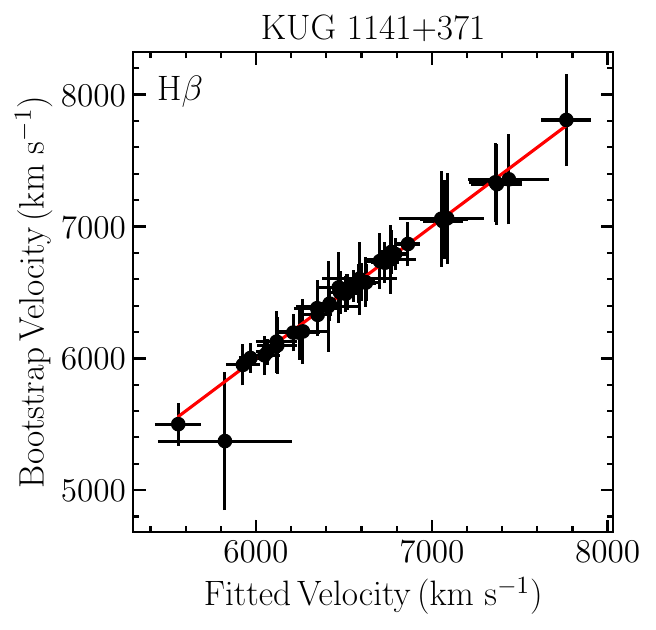}
\includegraphics[scale=0.32]{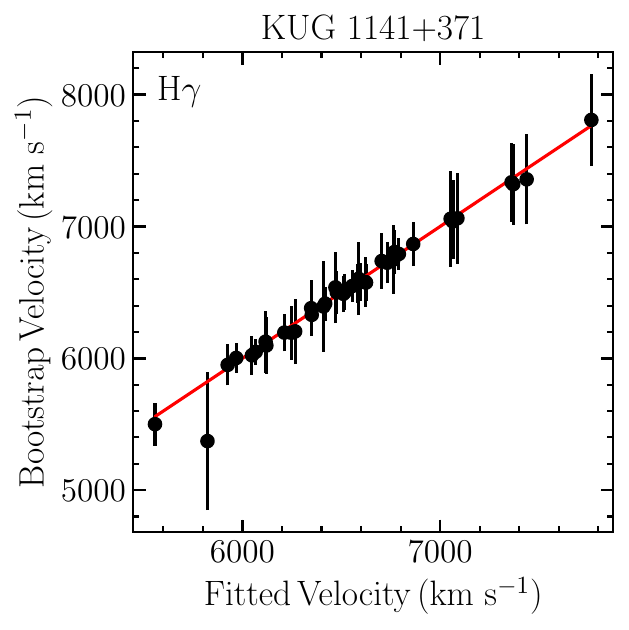}
\includegraphics[scale=0.32]{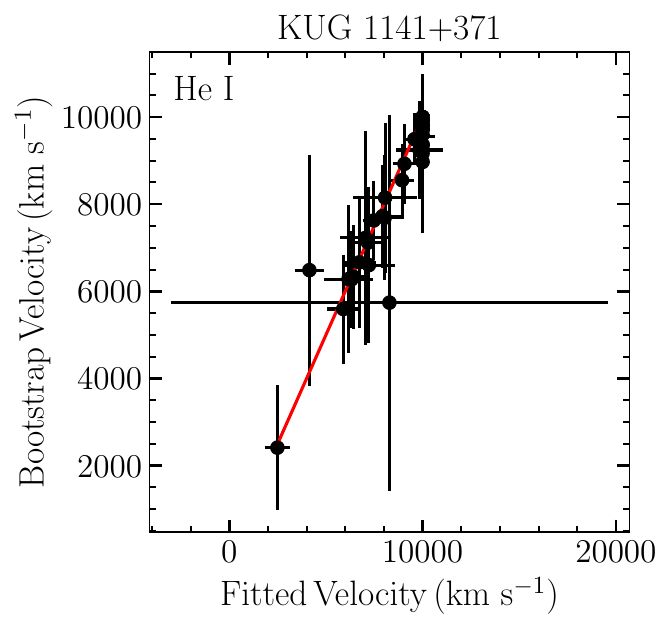}
\includegraphics[scale=0.32]{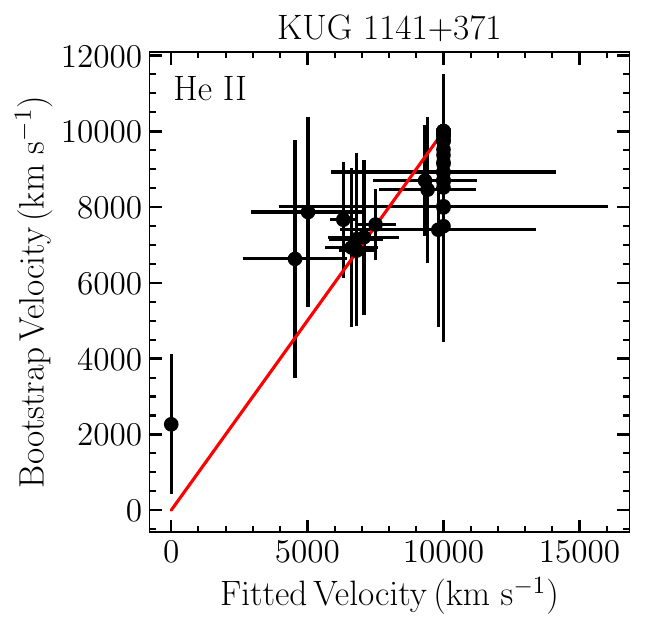}
\includegraphics[scale=0.32]{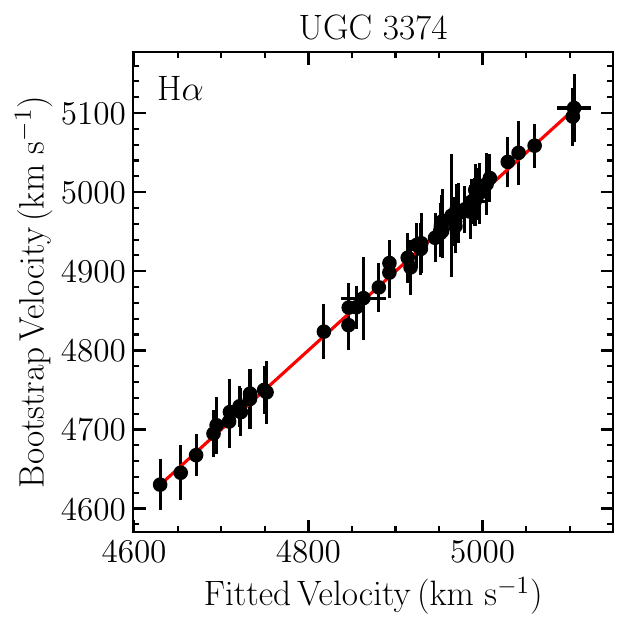}
\includegraphics[scale=0.32]{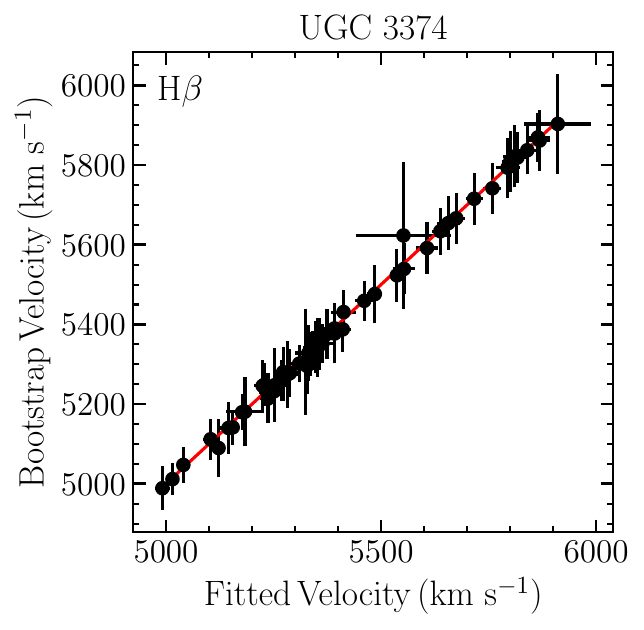}
\includegraphics[scale=0.32]{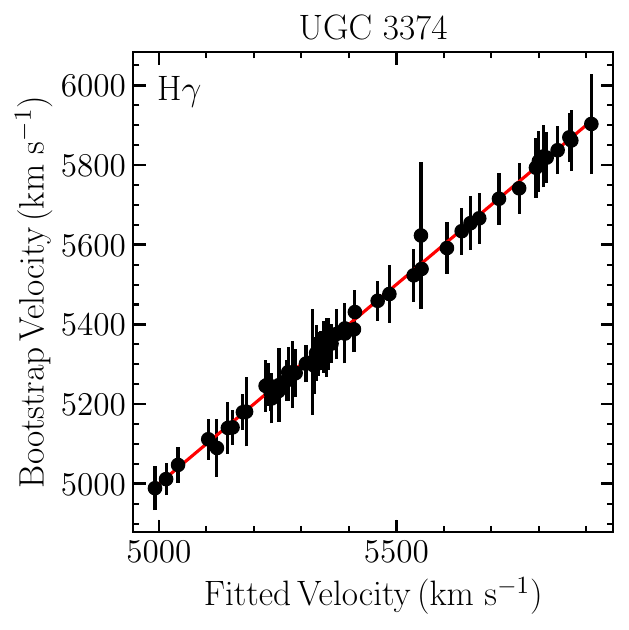}
\includegraphics[scale=0.32]{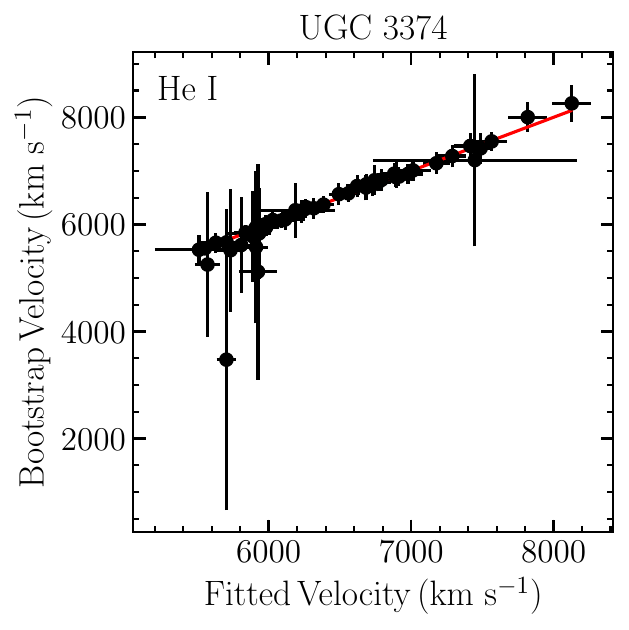}
\includegraphics[scale=0.32]{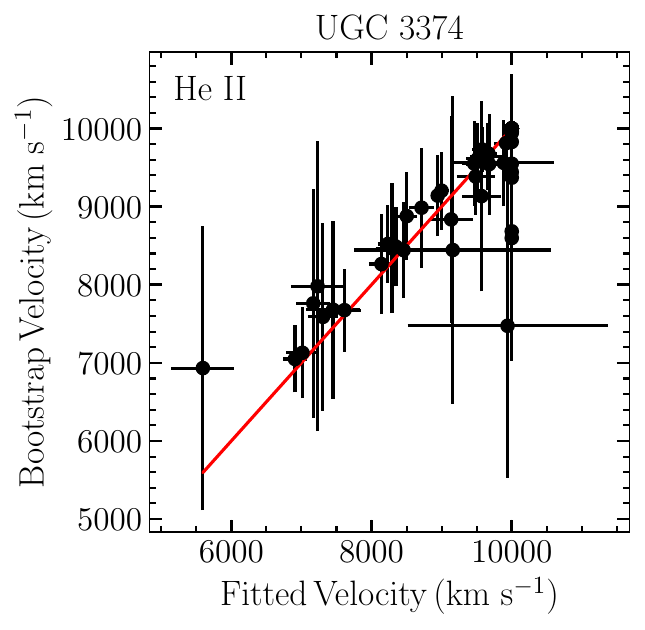}
\caption{Same as Figure \ref{fig:comparisonflux}, but showing the comparison of width measurements instead.
}
\label{fig:comparisonwidth}
\end{figure*}

Figures~\ref{fig:comparisonflux} and \ref{fig:comparisonwidth} present a comparison of the flux and FWHM measurements, respectively, obtained from our original fitting method and the Monte Carlo simulations. The results from both methods closely follow a 1:1 relation, validating the reliability of our spectral fitting measurements. In general, bootstrap resampling tends to yield larger uncertainties compared to other methods \citep{Peterson1998}. This effect is particularly significant for spectral fitting because the random subset selection process discards $\sim$37\% (1/$e$) of the data, corresponding to reducing the spectral resolution. For emission lines with low S/N, the uncertainties derived from Monte Carlo simulations are generally larger than those from our original fitting method. However, since the fitting results from both methods are essentially consistent, it is unlikely that the uncertainties in our original method are underestimated. For high-S/N lines, the uncertainties from the two methods are comparable. This consistency demonstrates the robustness of our original uncertainty estimates.

\end{document}